\documentclass[a4paper, 12pt]{article}  
\usepackage{pdfpages}
\usepackage{lmodern}
\usepackage{graphicx} 
\usepackage{wrapfig} 
\usepackage{textcomp} 
\usepackage[utf8]{inputenc}
\DeclareSymbolFont{letters}{OML}{ztmcm}{m}{it}
\usepackage{color}
\usepackage{bm}
\usepackage{bigdelim, multirow}

\usepackage{array}
\usepackage{listings}
\usepackage{adjustbox}
\usepackage[boxed]{algorithm2e}
\usepackage{subcaption}
\usepackage{multirow} 
\usepackage[OMLmathrm,OMLmathbf]{isomath}
\usepackage{booktabs} 
\usepackage{pdflscape}
\usepackage{natbib}
\usepackage{comment}

\usepackage[T1]{fontenc} 
\usepackage{geometry}
\usepackage{tablefootnote}
\usepackage{amsmath}
\usepackage{etoolbox} 
\usepackage{amsthm}
\usepackage{amsfonts}
\usepackage{amssymb}
\usepackage[section]{placeins}
\usepackage{diagbox} 
\usepackage{hyperref}
\usepackage{graphicx} 
\usepackage{fixmath} 
\usepackage[scaled=.90]{helvet}
\usepackage{wrapfig}
\usepackage{tikz}
\usepackage{footnote}
\makesavenoteenv{tabular}
\makesavenoteenv{table}
\usepackage{setspace}
\usepackage[boxed]{algorithm2e}
\usepackage[symbol]{footmisc}
\renewcommand{\thefootnote}{\arabic{footnote}}
\usepackage{multibib}
\usepackage{multicol}
\newcites{supp}{Supplementary Material References}

\usepackage{caption}
\usepackage{subcaption}

\usepackage[symbol]{footmisc}

\makeatletter
\renewcommand\@biblabel[1]{\textbf{#1.}} 
\usepackage{enumitem}

\usepackage{pifont}
\setlist{nolistsep,leftmargin=*}
\renewcommand{\maketitle}{ 
	\begin{center}
		{\LARGE\@title} 
		
		\vspace{0pt} 
		
		{\large\@author} 
		
		\vspace{30pt} 
	\end{center}
}

\renewcommand{\thefootnote}{\fnsymbol{footnote}} 

\title{Dependence matters: Statistical models to identify the drivers of tie formation in economic networks}
\author{\normalsize{Giacomo De Nicola$^{1,}\footnote[1]{Corresponding author:  giacomo.denicola@stat.uni-muenchen.de}$, Cornelius Fritz$^2$, Marius Mehrl$^3$, Göran Kauermann$^{1}$}}

\begin{document}

\maketitle
\begin{center}
    
\vspace{-1cm}
\normalsize{
Department of Statistics, LMU Munich, Ludwigstr. 33, 80539 Munich, Germany$^1$ \\
\vspace{0.1cm}
Department of Statistics, Pennsylvania State University, 100 Thomas Building, \\ 16802 State College, PA, USA$^2$ \\
\vspace{0.1cm}
School of Politics and International Studies, University of Leeds, LS2 9JT Leeds, \\ United Kingdom$^3$}

\end{center}

\begin{abstract}
Networks are ubiquitous in economic research on organizations, trade, and many other areas. However, while economic theory extensively considers networks, no general framework for their empirical modeling has yet emerged. We thus introduce two different statistical models for this purpose -- the Exponential Random Graph Model (ERGM) and the Additive and Multiplicative Effects network model (AME). Both model classes can account for network interdependencies between observations, but differ in how they do so. The ERGM allows one to explicitly specify and test the influence of particular network structures, making it a natural choice if one is substantively interested in estimating endogenous network effects. In contrast, AME captures these effects by introducing actor-specific latent variables affecting their propensity to form ties. This makes the latter a good choice if the researcher is interested in capturing the effect of exogenous covariates on tie formation without having a specific theory on the endogenous dependence structures at play. After introducing the two model classes, we showcase them through real-world applications to networks stemming from international arms trade and foreign exchange activity. We further provide full replication materials to facilitate the adoption of these methods in empirical economic research.\\

\noindent \textbf{Keywords} -- Inferential Network Analysis, Network Data, Endogeneity, Arms Trade, Foreign Exchange Networks, Statistical Modeling \\
\noindent \textbf{JEL classification} -- C20, C49, F14, F31, L14\\
\noindent {Declaration of interest: none}
\end{abstract}


\renewcommand{\thefootnote}{\arabic{footnote}}

\section{Introduction}
\label{sec:intro}
The study of networks has established itself as a central topic in economic research \citep{Jackson_2008}. Within the broader context of the study of complex and interdependent systems \citep[see e.g.][]{flaschel1997,flaschel2007,flaschel2018}, 
networks can be defined as interconnected structures which can naturally be represented through graphs. In the economic literature, networks have been extensively considered from a theoretical perspective, with the primary goal of understanding how economic behavior is shaped by interaction patterns \citep{jackson2007meeting}. Indeed, the adequate modelling of such interactions has been described as one of the main empirical challenges in economic network analysis \citep{Jackson_Rogers_Zenou_2017}. Research in this direction on, e.g., organizations as networks, diffusion in networks, network experiments, or network games, is surveyed in \citet{Bramoulle_Galeotti_Rogers_2016}, \citet{Jackson:2014}, and \citet{Jackson_Rogers_Zenou_2017}. These theoretical advances find application in many different fields in which network structures naturally arise, such as national and international trade, commercial agreements, firms' organization, and collaboration activity.  
However, such advances have not yet been accompanied by a corresponding shift in the standard methods used to empirically validate them. Some recent contributions \citep[see e.g.][]{atalay2011network,Chaney_2014,Morales_Sheu_Zahler_2019} develop estimators tailored specifically to their network-based theoretical models, but more generally applicable modeling frameworks for the analysis of real-world network data have not yet emerged. Statistical methods specifically designed to empirically test theories where interdependencies arise from network structures, such as the Exponential Random Graph Model (ERGM), exist but are not yet widely used by economists.
\citet{Jackson:2014}, for instance, discusses ERGMs but argues that they ``suffer from proven computational problems'' (\citeyear[p.76]{Jackson:2014}). \citet{Jackson_Rogers_Zenou_2017} further explain that ``it is practically impossible to estimate the likelihood of a given network at even a moderately large scale'', concluding that with ERGMs, ``there is an important computational hurdle that must be overcome in working with data''  (\citeyear[p.85]{Jackson_Rogers_Zenou_2017}). 



Contrasting this assessment, we argue that recent work in the realm of empirical network analysis provides robust and scalable methods with readily available implementations in the $\mathtt{R}$ statistical software \citep{R}. Computational issues thus do not represent an insurmountable barrier to employ robust inferential network methods anymore. In this paper, we demonstrate the effectiveness and usability of some of those methods by applying them to real economic data. We specifically focus on models which aim to capture the mechanisms leading to network formation, i.e. to measure how the probability of forming a tie is influenced by (a) nodal characteristics, (b) pairwise covariates, and (c) the rest of the network.  In particular, our focus is on Exponential Random Graph Models (ERGM) \citep{robins_introduction_2007} and Additive and Multiplicative Effect (AME) network models \citep{hoff2021}, respectively implemented in the  $\mathtt{R}$ packages $\mathtt{statnet}$ \citep{handcock2008} and $\mathtt{amen}$ \citep{hoff2015}. 
We find these two model classes to be among the most promising ones for applications in the economic sciences, as they are well suited for answering two broad categories of research questions. The ERGM is an ideal fit if, based on economic theory, the researcher envisages a particular dependence structure for the existence of ties in the network at hand and wants to test whether their theory is corroborated by empirical data. On the other hand, AME, and more generally continuous latent variable models, are a good choice when the researcher is interested in capturing the effect of exogenous variables on tie formation without having prior knowledge on which endogenous network dependence mechanisms are at play. In this case, AME offers the possibility to estimate the effect of both nodal and pairwise covariates while simultaneously controlling for network effects, which may induce bias if ignored (see \citealp{lee_network_2021}). In addition, the estimated latent structure can provide insight on the underlying network mechanisms for which they are controlling.

The principal aim of this paper is to showcase ERGM and AME by focusing on their value for economic research. After introducing each model class, we demonstrate their empirical usage by respectively applying them to two relevant economic questions stemming from real-world networks. We first use the ERGM to model the international trade of major conventional weapons, where a directed tie exists if one country transfers arms to another. In line with \citet{Chaney_2014}, network effects such as directed triadic closure (e.g. the positive impact of an increase in the volume of trade between countries A and B on the probability that country C, that already exports to A, starts exporting
to B) are of explicit theoretical interest in this application, and the ERGM allows for their proper specification and testing. We then make use of the AME model to study a historical network of global foreign exchange activity, where a directed edge is present if one country's national currency is actively traded within the other country. AME allows us to estimate how relevant country features, such as per-capita gdp and the gold standard, and pairwise covariates, such as the distance between two countries and their reciprocal trade volume, influence tie formation, while controlling for network effects to provide unbiased estimates. We further compare the two model classes, weighing pros and cons of each approach and providing guidance on which tool is appropriate for applications to different empirical settings and research questions. %
Finally, in addition to a step-by-step analysis and interpretation of these application cases, we  provide full replication code in our GitHub repository\footnote{https://github.com/gdenicola/statistical-network-analysis-in-economics}, allowing for seamless reproducibility. We, therefore, demonstrate the ``off-the-shelf'' applicability of these methods, and offer applied researchers a head-start in employing them to study substantive economic problems.


Our contribution is related to various strands of the growing  literature on economic networks (e.g.\ \citealp{jackson2007meeting, Jackson_2008,Bramoulle_Galeotti_Rogers_2016}). Due to its focus on economic questions, our work differs from surveys in physics \citep{newman2003}, statistics \citep{goldenberg2010}, or political science \citep{cranmer2017navigating}. Several articles provide overviews and surveys of existing economic network models from a theoretical perspective \citep{Jackson:2014, graham2015methods, Jackson_Rogers_Zenou_2017, Paula_2020}. None of these articles concentrates on discussing broadly applicable statistical modeling frameworks, such as ERGM and AME, from an empirical perspective. 
In this sense our paper is similar in spirit to \citet{Pol_2019} who, however, only focuses on ERGM, without comparing alternative approaches. Indeed, one of the goals of this paper is to shed light on the emerging AME model class (and, more generally, on latent variable network models) for future applications in the economic literature.

The remainder of the paper is structured as follows. 
Section 2 discusses existing literature and presents the mathematical and notational framework used to define and discuss networks throughout the paper. Section 3 introduces the ERGM and applies it to the international arms trade network. Section 4 is dedicated to AME and its application to the global foreign exchange network. Section 5 concludes the paper with a brief discussion on the two model classes, contrasting their different uses and highlighting pros and cons of each approach.


\section{Economic Networks}
\label{sec:litrev}

\subsection{Related literature}
Even though network structures naturally arise in many aspects of economics and are subject of prominent research in the field, much of the previous literature has ignored the implied interdependencies, instead opting for regression models assuming ties to be independent conditional on the covariates (e.g.\ \citealp{anderson2003}, \citealp{rose2004}, \citealp{lewer2008}). This assumption is often unreasonable in practice. I
t would, for example, imply that Germany imposing economic sanctions on Russia is independent of Italy imposing sanctions on Russia, and, in the directed case, even of Russia imposing them on Germany itself. While no standard framework for the modeling of empirical network data has emerged in economics so far, a number of contributions in -- or adjacent to -- the field do make use of statistical network models. We shortly survey these works here to show that the models we present are indeed suitable for the analysis of economic data. Possibly the most obvious kind of economic network is the international trade network \citep[see][]{Chaney_2014} and many of these studies accordingly seek to model the formation of trade ties. In this vein, two early studies \citep{Ward_Hoff_2007,Ward_Ahlquist_Rozenas_2013} apply latent position models to show that trade exhibits a latent network structure beyond what a standard gravity model can capture \citep[see also][]{Fagiolo_2010,Duenas_Fagiolo_2013}. More recently, numerous contributions have used the ERGM to explicitly theorize and understand network interdependence in the general trade \citep{Herman_2022,Liu_Shen_Sun_Yan_Hu_2022,Smith_Sarabi_2022} as well as the trade in arms \citep{Thurner_Schmid_Cranmer_Kauermann_2019,Lebacher_Thurner_Kauermann_2021a}, patents \citep{He_Dong_Wu_Jiang_Zheng_2019}, and services \citep{Feng_Xu_Wu_Zhang_2021}.

That being said, empirical research on economic networks is not limited to trade. \citet{Smith_Gorgoni_Cronin_2019} use multilevel ERGMs to study a production network consisting of ownership ties between firms at the micro-level and trade ties between countries at the macro-level, while \citet{Mundt_2021} explores the European Union's sector-level production network via ERGMs as well as an alternative methodology, the stochastic actor-oriented model (SAOM). The latter is another prominent tool in the realm of network analysis, which is suitable for modeling longitudinal network data. As we, in the interest of brevity, focus on models for static networks (i.e. networks that are observed only at one point in time), we do not treat the SAOM, and instead refer to \citet{snijders1996, snijders2017} for an introduction to the model class. Going back to empirical research on economic networks in the literature, \citet{fritz2022modelling} deploy ERGMs to investigate patent collaboration networks. Studies on foreign direct investments document network influences using latent position models \citep{Cao_Ward_2014}, or seek to model them via extensions of the ERGM \citep{Schoeneman_Zhu_Desmarais_2022}. Finally, economists also study networks of interstate alliances and armed conflict \citep[see e.g.][]{Jackson_Nei_2015,Konig_Rohner_Thoenig_Zilibotti_2017}, both of which have been modeled via ERGMs \citep{Cranmer_Desmarais_Kirkland_2012,Campbell_Cranmer_Desmarais_2018} and AME \citep{Dorff_Gallop_Minhas_2020,Minhas_Dorff_Gallop_Foster_Liu_Tellez_Ward_2021}. This short survey indicates that both ERGM and AME can be used to answer questions which are of substantive interest to economists. 


\subsection{Setup}
Before introducing models for networks in which dependencies between ties are expected, we briefly introduce the mathematical framework for networks, as well as the necessary notation. Let $\bm{y} = \left(y_{ij}\right)_{i,j = 1, ..., n}$ be the adjacency matrix representing the observed binary network, comprising $n$ fixed and known agents (nodes). In this context, $y_{ij} = 1$ indicates an edge from agent $i$ to agent $j$, while $y_{ij} = 0$ translates to no edge between the two. Since self-loops are not admitted for most studied networks, the diagonal of $\bm{y}$ is left unspecified or set to zero. Depending on the application, the direction of an edge can carry additional information. If it does, we call the network directed. In this article, we mainly focus on this type of networks. Also note that all matrix-valued objects are written in bold font for consistency. In addition to the network connections, we often observe covariate information on the agents, which can be at the level of single agents (e.g.\ the gdp of a country) or at the pairwise level (e.g.\ the distance between two countries). We denote covariates by $\bm{x}_1, ..., \bm{x}_p$, 
 and our goal is to specify a statistical model for $\bm{Y}$, that is the random variable corresponding to $\bm{y}$, conditional on $\bm{x}_1, ..., \bm{x}_p$. 
A natural way to do this is to specify a probability distribution over the space of all possible networks, which we define by the set $\mathcal {Y} $. Two main characteristics differentiate our modeling endeavor from classical regression techniques, such as Probit or logistic regression models. First, for most applications, we only observe one realization $\bm{y}$ from $\bm{Y}$, rendering the estimation of the parameters to characterize this distribution particularly challenging. Second, the entries of $\bm{Y}$ are generally co-dependent; thus, most conditional dependence assumptions inherent to common regression models are violated. 
Generally, we term mechanisms that induce direct dependence between edges to be endogenous, while all effects external to the modeled network, such as covariates, are called exogenous.

\section{The Exponential Random Graph Model}
\label{sec:ergm}

The ERGM is one of the most popular models for analyzing network data. First introduced by \citet{holland1981} as a model class that builds on the platform of exponential families, it was later extended with respect to fitting algorithms and more complex dependence structures \citep{lusher_exponential_2012, robins_recent_2007}. We next introduce the model step-by-step to highlight its ability to progressively generalize by building on conditional dependence assumptions. 

\subsection{Accounting for dependence in networks}
We begin with the simplest possible stochastic network model, the Erdös-Rényi-Gilbert model \citep{erdos_random_1959,gilbert_random_1959}, 
 where all edges are assumed to be independent and to have the same probability of being observed. In stochastic terms, each observed tie is then a realization of a binomial random variable with success probability $\pi$, which yields
\begin{align}
    \mathbb{P}_\pi (\bm{Y} = \bm{y}) =  \prod_{i = 1}^n \prod_{j \neq i}  \pi^{y_{ij}} (1-\pi)^{1-y_{ij}}
    \label{eq:bernoulli_graph}
\end{align}
for the probability to observe $\bm{y}$. Evidently, model \eqref{eq:bernoulli_graph}, which implies equal probability for all possible ties, is too restrictive to be applied to real world problems. In the next step, we, therefore, additionally incorporate covariates $x_{ij}$ by letting $\pi$ vary depending on those covariates, leading to edge-specific probabilities $\pi_{ij}$. Following the common practice in logistic regression, we parameterize the log-odds by $\log\left(\frac{\pi_{ij}}{1-\pi_{ij}}\right) = \theta^\top x_{ij}$, where $x_{ij}$ is a vector of exogenous statistics with the first entry set to 1 to incorporate an intercept, and get
\begin{align}
    \mathbb{P}_\theta (\bm{Y} = \bm{y}) =  \prod_{i = 1}^n \prod_{j \neq i}  \left(\frac{\exp\{\theta^\top x_{ij}\}}{1+\exp\{\theta^\top x_{ij}\}}\right)^{y_{ij}} \left(\frac{1}{1+\exp\{\theta^\top x_{ij}\}}\right)^{1-y_{ij}}.
    \label{eq:bernoulli_graph_cov}
\end{align}
From \eqref{eq:bernoulli_graph_cov}, the analogy to standard logistic regression being a special case of generalized linear models \citep{nelder_generalized_1972} becomes apparent. The joint distribution of $\bm{Y}$ can be formulated in exponential family form, yielding
\begin{align}
    \mathbb{P}_{\theta} (\bm{Y} = \bm{y}| x) =  \frac{\exp\{\theta^\top s(\bm{y})\}}{\kappa(\theta)},
    \label{eq:bernoulli_graph_factor}
\end{align}
where $s(\bm{y}) = (s_1(\bm{y}), ..., s_p(\bm{y}))$, $s_q(\bm{y}) = \sum_{i = 1}^n  \sum_{j \neq i} y_{ij}x_{ij,q} ~ \forall ~q = 1, ..., p$, with $x_{ij,q}$ as $q-th$ entry in $x_{ij}$ and $\kappa(\theta) = \prod_{i = 1}^n \prod_{j \neq i}(1 + \exp\{\theta^\top x_{ij}\})$. In the jargon of exponential families, we term $s(\bm{y})$ sufficient statistics.  

\citet{newcomb_reciprocity_1979} observed that many observed networks exhibit complicated relational mechanisms, including reciprocity, which we can account for by extending the set of sufficient statistics. Under reciprocity, an edge $Y_{ji}$ influences the probability of its reciprocal edge $Y_{ij}$ to occur. Analyzing social networks, we would expect that the probability of agent $i$ nominating agent $j$ to be a friend is higher if agent $j$ has nominated agent $i$ as a friend. \citet{holland1981} extended model \eqref{eq:bernoulli_graph} to such settings with the so-called $p_1$ model. To represent reciprocity, we assume dyads, each of them defined by $(Y_{ij}, Y_{ji})$, to be independent of one another, which again yields an exponential family distribution similar to \eqref{eq:bernoulli_graph_factor} with sufficient statistics 
 that count the number of mutual ties ($s_{\text{Mut}}(\bm{y}) = \sum_{i<j} y_{ij}y_{ji}$), of edges ($s_{\text{Edges}}(\bm{y}) =  \sum_{i = 1}^n  \sum_{j \neq i} y_{ij}$), and the in- and out-degree statistics for all  degrees observed in the networks\footnote{We provide more details on this derivation in the Supplementary Material.}. Agents' in- and out-degrees are their number of incoming and outgoing edges, and relate to their relative position in the network \citep{wasserman_social_1994}.  

\begin{figure}
    \centering\captionsetup[sub]{font=large}
    \begin{subfigure}[c]{0.24\textwidth}
    \includegraphics[width=\textwidth, page = 1]{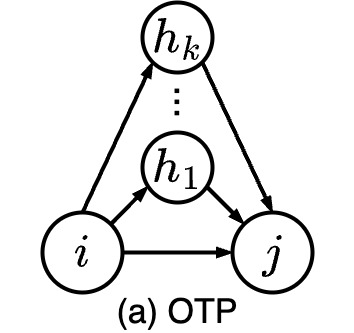}
    \label{fig:otp}
    \end{subfigure}
    \begin{subfigure}[c]{0.24\textwidth}
    \includegraphics[width=\textwidth, page = 1]{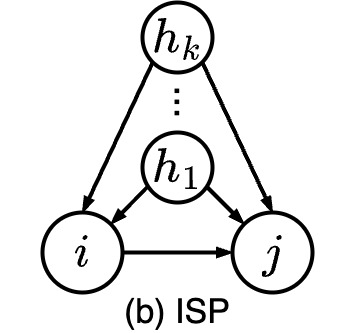}
    \label{fig:isp}
    \end{subfigure}
      \begin{subfigure}[c]{0.24\textwidth}
    \includegraphics[width=\textwidth, page = 1]{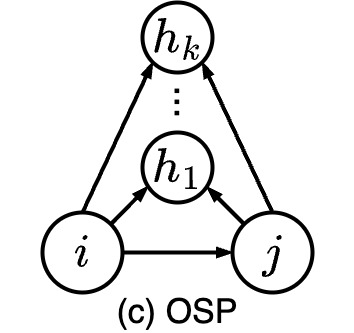}
    \label{fig:osp}
    \end{subfigure}
    \begin{subfigure}[c]{0.24\textwidth}
    \includegraphics[width=\textwidth, page = 1]{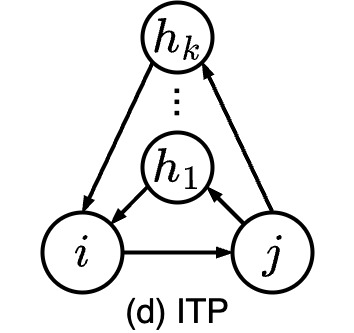}
    \label{fig:itp}
    \end{subfigure}
    \caption{Illustration of directed edgewise-shared partner statistics for $k$ agents. Circles represent agents, and black lines represent edges between them. The names follow $\mathtt{statnet}$ nomenclature: OTP = ``Outgoing Two-Path'', ISP = ``Incoming Shared Partner'', OSP = ``Outgoing Shared Partner'', and ITP = ``Incoming Two-Path''.}
    \label{fig:triangular_statistics}
\end{figure}

Next to reciprocity, another important endogenous network mechanism is transitivity, originating in the structural balance theory of \citet{heider_attitudes_1946} and adapted to binary networks by \citet{davis_clustering_1970}. Transitivity affects the clustering in the network, implying that a two-path between agents $i$ and $j$, i.e. $y_{ih} = y_{hj} = 1$ for some other agent $h$, affects the edge probability of $Y_{ij}$. Put differently, $Y_{ij}$ and $Y_{kh}$ are assumed to be independent iff $i,j \neq k$ and $i,j \neq h$. \citet{frank_markov_1986} proposed the Markov model to capture such dependencies. For this model, the sufficient statistics are star-statistics, which are counts of sub-structures in the network where one agent has (incoming and outgoing) edges to between 0 and $n-1$ other agents, and counts of triangular structures. If the network is directed it is possible to define different types of triangular structures, as depicted in Figure \ref{fig:triangular_statistics}.   

\subsection{Extension to general dependencies}
Starting from the Erdös-Rényi-Gilbert model, which is a special case of a generalized linear model, we have consecutively allowed for more complicated dependencies between edges, resulting in the Markov graphs of \citet{frank_markov_1986}. Over this course, we showed that each model can be stated in exponential family form, characterized by a particular set of sufficient statistics. We now make this more explicit to allow for more general dependence structures, and specify a probabilistic model for $\bm{Y}$ directly through the sufficient statistics\footnote{Alternatively, \eqref{eq:ergm_base} can also be derived as the equilibrium distribution of a strategic game where players myopically reassess and update their links to optimize their utility in the network \citep[see][]{mele_structural_2017,boucher_my_2017}.}.    
\citet{wasserman_logit_1996} introduced this model as
\begin{align}
    \label{eq:ergm_base}
   \mathbb{P}_{\theta}(\mathbf{Y} = \mathbf{y}) = \frac{\exp\{\theta^\top \boldsymbol{s}(\mathbf{y})\}}{\kappa(\theta)}, 
\end{align}
where 
$\theta$ is a $p$-dimensional vector of parameters to be estimated, $\boldsymbol{s}(\mathbf{y})$ is a function calculating the vector of $p$ sufficient statistics for network $\mathbf{y}$, 
and $\kappa(\theta) = \sum_{\Tilde{\mathbf{y}} \in \mathcal{Y}} \exp\{\theta^\top s(\Tilde{\mathbf{y}})\}$ is a normalizing constant to ensure that \eqref{eq:ergm_base} sums up to one over all $\mathbf{y} \in \mathcal{Y}$.
To estimate $\theta$, \citet{handcock_assessing_2003} adapted the Monte Carlo Maximum Likelihood technique  of \citet{geyer_constrained_1992}, approximating the logarithmic likelihood ratio of $\theta$ and a fixed $\theta_0$ via Monte Carlo quadrature (see \citealp{hunter_computational_2012}, for an in-depth discussion). 

A problem often encountered when fitting model \eqref{eq:ergm_base} to networks is degeneracy \citep{handcock_assessing_2003,schweinberger_instability_2011}. Degenerate models are characterized by probability distributions that put most probability mass either on the empty or on the full network, i.e., where either all or no ties are observed. To detect this behavior, one can use a goodness-of-fit procedure where observed network statistics are compared to statistics of networks simulated under the estimated model \citep{hunter_goodreau_handcock_2008}. To address it, \citet{snijders_new_2006} and \citet{hunter_inference_2006} propose weighted statistics that, in many cases, have better empirical behavior. Degeneracy commonly affects model specifications encompassing statistics for triad counts and multiple degree statistics. For in-degree statistics, we would thus incorporate the geometrically-weighted in-degree,
\begin{align}
    \label{eq:stat_gwdeg}
    GWIDEG(\mathbf{y},\alpha) = \exp\{\alpha\} \sum_{k = 1}^{n-1} \left(1-\left(1-\exp\{-\alpha\}\right)^k\right) IDEG_k(\mathbf{y}),
\end{align}
where $IDEG_k(\mathbf{y})$ is the number of agents in the studied network with in-degree $k$ and $\alpha$ is a fixed decay parameter. One can substitute $IDEG_k(\mathbf{y})$ in \eqref{eq:stat_gwdeg} with the number of agents with a specific out-degree, $ODEG_k(\mathbf{y})$, to capture the out-degree distribution. We term these statistics geometrically weighted since the weights in \eqref{eq:stat_gwdeg} are a geometric series\footnote{Geometrically weighted statistics require setting the decay parameter $\alpha$. We set $\alpha = \log(2)$, though it can also be estimated as an additional parameter given sufficient data \citep{hunter_inference_2006}.}. A positive estimate implies that an edge from a low-degree agent is more likely than an edge from a high-degree agent, resulting in a decentralized network. If, on the other hand, the corresponding coefficient is negative, one may interpret it as an indicator for a centralized network. 

To capture clustering, we have to define the distribution of edgewise-shared partners (ESP). This distribution is defined as the relative frequency of edges in the network with a specific number of $k$ shared partners, that we denote by $ESP(\mathbf{y})$ for $k \in \{1, ..., n-2\}$. 
As shown in Figure \ref{fig:triangular_statistics}, various versions of edgewise-shared partner statistics can be found in directed networks, depending on the direction of the edges between the three  agents involved. Geometrically weighted statistics can be stated for them in a similar manner as for degree statistics. For example, for the outgoing two-path (OTP, see Figure \ref{fig:triangular_statistics}a), this is 
\begin{align}
    \label{eq:stat_gwotp}
    GWOTP(\mathbf{y},\alpha) = \exp\{\alpha\} \sum_{k = 1}^{n-2} \left(1-\left(1-\exp\{-\alpha\}\right)^k\right) OTP_k(\mathbf{y}).
\end{align}
In this case, a positive coefficient indicates that sharing ties with third actors increases the probability of observing an event between two agents.      

Along with capturing endogenous network statistics, it is also possible to extend the ERGM framework to include the temporal dimension, that is, to model longitudinal network data. This is done quite naturally through use of a Markov assumption on the temporal dependence of subsequently observed networks, giving rise to the Temporal Exponential Random Graph Model (TERGM). As we here focus on static networks, we do not cover this in depth, and refer to \citet{Hanneke_Fu_Xing_2010} for an introduction to the TERGM, and to \cite{Fritz2019} for a more general discussion on temporal extensions to the model class. 

In summary, the ERGM allows to account for network dependencies via explicitly specifying them in $\boldsymbol{s}(\mathbf{y})$. A large variety of potential network statistics, such as those given in \eqref{eq:stat_gwdeg} and \eqref{eq:stat_gwotp}, can be included in $\boldsymbol{s}(\mathbf{y})$, enabling to test for their influence in the formation of the observed network. By allowing for this explicit inclusion and testing of network statistics, the ERGM requires researchers to at least have an implicit theory regarding what types of network dependence should exist in the network they study. Without such theory to guide the selection of network statistics, the range of potential network dependencies, and corresponding statistics, is virtually endless\footnote{For a survey of possible endogenous terms, see \citet{morris_specification_2008}.}. As a result, the ERGM is best suited for research questions that explicitly concern interdependencies within the network. If these interdependencies are, instead, only a potential source of bias the researcher wants to control for, the AME model (introduced in Section \ref{sec:amen}) may be a better fit.         


\subsection{Application to the international arms trade network} 

\begin{figure}[t]
\centering 
\includegraphics[width=0.85\linewidth, trim=0.6cm 8cm 0.6cm 8cm, clip]{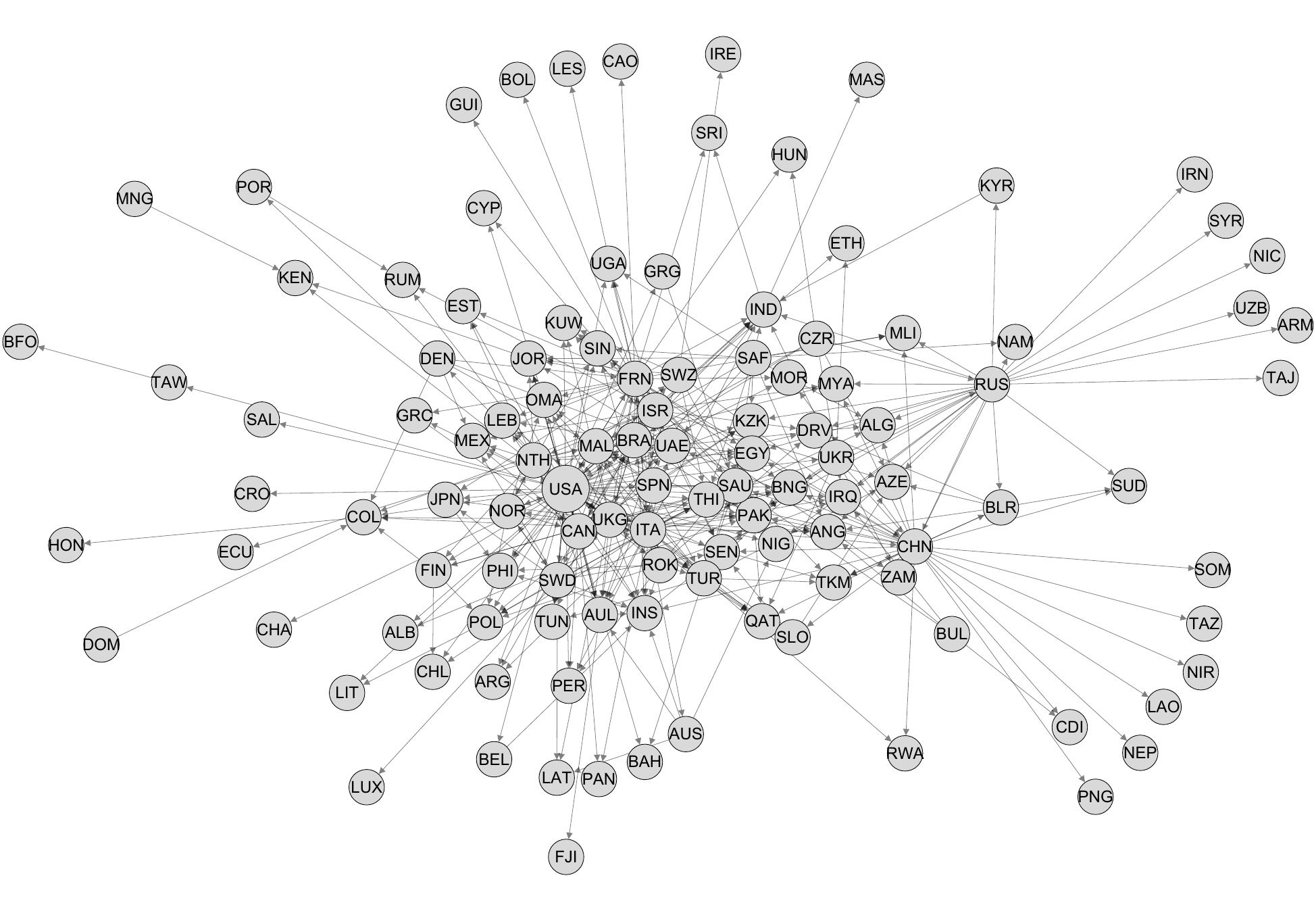}
\caption{Illustration of the international arms trade network in 2018. Countries are labeled by their ISO 3166-1 codes, and a directed edge from node $i$ to node $j$ indicates major conventional weapons being delivered from country $i$ to country $j$.}\label{fig:mcwnet}
\end{figure}

We next make use of the ERGM to analyze the international arms transfer network. Recent studies on trade in Major Conventional Weapons (MCW), such as fighter aircraft or tanks, not only emphasize its networked nature, but also argue that this very nature is of substantive theoretical interest \citep{Thurner_Schmid_Cranmer_Kauermann_2019,Fritz_Thurner_Kauermann_2021}. In line with \citet{Chaney_2014}, triadic trade structures are held to reveal information regarding the participants' economic and security interests. Explicitly modeling these structures allows us to test hypotheses regarding their effects on further arms transfers. Accordingly, we seek to model the network of international arms transfers in the year 2018, where countries are nodes and a directed edge indicates MCW being delivered from country $i$ to country $j$.  Our interest here mainly lies in uncovering the network's endogenous mechanisms.
MCW trade data come from \citet{SIPRI_2021}, and the resulting network is depicted in Figure \ref{fig:mcwnet}, obtained using the Yifan Hu force-directed graph drawing algorithm \citep{hu2005} with the software Gephi \citep{bastian_gephi_2009}.

For estimating the parameters characterizing the ERGM, we use the \texttt{R} package $\mathtt{ergm}$ \citep{handcock_ergm_2008}. Since evaluating $\kappa(\bm{\theta})$ from \eqref{eq:ergm_base} necessitates calculating the sum of $|\mathcal{Y}| = 2^{n(n-1)}$ terms, we rely on MCMC approximations thereof to obtain the maximum likelihood estimates (see \citealp{handcock_assessing_2003} and \citealp{hummel_improving_2012} for additional information on this topic). As discussed above, the ERGM allows us to use both exogenous (node-specific and pair-specific) attributes as well as endogenous structures to model the network of interest. Here, we select both types of covariates based on existing studies on the arms trade \citep{Thurner_Schmid_Cranmer_Kauermann_2019,Fritz_Thurner_Kauermann_2021}. In addition to an edges term, which corresponds to the intercept in standard regression models, we include importers' and exporters' logged GDP, whether they share a defense pact, their absolute difference in ``polity'' scores (a type of democracy index), and their geographical distance\footnote{Data for these covariates come from the $\mathtt{peacesciencer}$ package \citep{Miller_2022}.}. We lag these covariates by three years, reflecting the median time between order and delivery for MCW delivered in 2018\footnote{We use the median as the distribution of times between order and delivery is quite skewed. As shown in the Supplementary Materials, our substantive results remain unchanged when using 4- and 5-year lags instead, which reflect the average time between order and delivery. In particular, the ERGM outperforms the logistic regression model regardless of lag choice.}. More importantly, for the purpose of demonstrating how to model network data with the ERGM, we specify five endogenous network terms. In- and out-degree (IDEG and ODEG) measure, respectively, importers' and exporters' trade activity, and thus capture whether highly active importers and exporters are particularly attractive trading partners, or if they are instead less likely to form additional trade ties. Moreover, we specify a reciprocity term to capture whether countries tend to trade MCW uni- or bidirectionally. We further include two types of triadic structures, which represent transitivity and a shared supplier between countries $i$ and $j$. The transitivity term counts how often country $i$ exports arms to $j$ while $i$ exports to $k$, which in turn exports to $j$, thus capturing $i$'s tendency to directly trade with $j$ if they engage in indirect trade (OTP, see Figure \ref{fig:triangular_statistics}a). In contrast, the shared supplier term counts how often country $i$ sends arms to $j$ while both import weapons from a shared supplier $k$ (ISP, see Figure \ref{fig:triangular_statistics}b). Note that, given the issue of degeneracy discussed above, we use geometrically weighted versions of all endogenous statistics except reciprocity. Finally, we include a repetition term capturing whether arms transfer dyads observed in 2018 had already occurred in any of the three previous years.
\begin{table}
\begin{center}
\begin{tabular}{l c c}
\hline
 & ERGM & Logit \\
\hline
Intercept                             & $-15.356 \; (2.017)^{***}$ & $-28.197 \; (1.731)^{***}$ \\
Repetition                      & $3.254 \; (0.141)^{***}$   & $3.957 \; (0.141)^{***}$   \\
Distance                   & $-0.081 \; (0.087)$        & $-0.239 \; (0.088)^{**}$   \\
Abs. Diff. Polity                    & $-0.001 \; (0.010)$        & $-0.003 \; (0.012)$        \\
Alliance  & $0.350 \; (0.207)$         & $0.209 \; (0.207)$         \\
log-GDP (Sender)                    & $0.300 \; (0.050)^{***}$   & $0.588 \; (0.045)^{***}$   \\
log-GDP (Receiver)                    & $0.166 \; (0.049)^{***}$   & $0.355 \; (0.039)^{***}$   \\
Mutual                            & $-0.311 \; (0.438)$        &                            \\
GWIDEG    & $-1.478 \; (0.296)^{***}$  &                            \\
GWODEG    & $-2.848 \; (0.296)^{***}$  &                            \\
GWOTP & $-0.146 \; (0.104)$        &                            \\
GWISP & $0.210 \; (0.083)^{*}$     &                            \\
\hline
AIC                               & $1769.718$                 & $1891.984$                 \\
BIC                               & $1866.053$                 & $1948.179$                 \\
Log Likelihood                    & $-872.859$                 & $-938.992$                 \\
\hline
\multicolumn{3}{l}{\scriptsize{$^{***}p<0.001$; $^{**}p<0.01$; $^{*}p<0.05$}}
\end{tabular}
\caption{Estimated coefficients and  standard errors (in parentheses) of the ERGM and the logistic regression model for the international arms trade network in 2018.}
\label{table:coefficients}
\end{center}
\end{table}
Results of this ERGM, as well as, for comparison's sake, a logistic regression that includes the same exogenous covariates but does not capture any of the endogenous network structures, are presented in Table \ref{table:coefficients}. These results can be compared directly, as, just like in a logistic regression model, coefficients in the ERGM indicate the additive change in the log odds of a tie occurring in association with a unit change in the respective variable. In this sense, the logistic regression model can be viewed as a special case of the ERGM in which the network effects are omitted. From the table, we can see how the two models differ both in their in-sample fit, as captured by AIC and BIC\footnote{As shown in the Supplementary Material, the ERGM also outperforms the Logit model when assessing their respective areas under the receiver-operator and precision-recall curves. In line with \citet{hunter_inference_2006}, one could also calculate the likelihood ratio test statistic from the log likelihoods reported in Table \ref{table:coefficients} for the same purpose.}, as well as in the substantive effects they identify for the exogenous covariates. The repetition coefficient is positive and statistically significant in both models, but differs substantially in its size. An arms transfer edge having occurred at least once in 2015-17 increases the log odds of it occurring also in 2018 by 3.96 in the Logit, but only by 3.26 in the ERGM. Similarly, both models agree that the log odds of an arms transfer occurring increase with the economic size of the sender and receiver, as captured by their respective GDPs, but the coefficients retrieved by the Logit are approximately double the size of those in the ERGM, thus attributing more explanatory power to them. Also in this vein, the effect of the geographical distance between sender and receiver is three times as large in the logistic regression as in the ERGM and, while statistically significant in the former, indistinguishable from zero in the latter. Finally, both models report small and statistically insignificant effects for countries' polity difference and alliance ties. Taken together, however, there are clear, substantively meaningful differences in the effect sizes and, in the case of geographical distance, even statistical significance of the coefficients that the ERGM and Logit recover for the exogenous covariates. 

Furthermore, three of the endogenous statistics included in the ERGM exhibit statistically significant effects on the probability of arms being traded. The results for in- and out-degree replicate the finding by \cite{Thurner_Schmid_Cranmer_Kauermann_2019}, showing that highly active importers and exporters are less likely to form additional trade ties. 
In the ERGM, coefficients can also be interpreted at the global level, in addition to the edge-level interpretation given above. The shared supplier term having a (statistically significant) positive coefficient indicates, at the edge level, that an exporter is more likely to transfer weapons to a potential receiver if both of them import arms from the same source. Globally, on the other hand, the same coefficient means that the observed network exhibits more shared supplier configurations -- where country $i$ sends weapon to $j$ while both receive arms from $k$ -- than would be expected in a random network of the same size. On the whole, the results presented in Table \ref{table:coefficients} offer an example for the striking differences that modeling network structures (instead of assuming them away) can make. The ERGM and Logit, while identical in their non-network covariates, report substantively different effects for these covariates, and, in the ERGM, network effects are also found to drive the formation of arms transfer edges.  

\section{The Additive and Multiplicative Effect Network Model}
\label{sec:amen}

\subsection{Latent variable network models}

Another way to account for network dependencies is by making use of latent variables. Models within this class assume that latent variables $Z_i$ are associated with each node $i$. Depending on the type of model, these latent variables can either be discrete (e.g.\ indicating group memberships for each node) or continuous, and affect the connection probability in different ways \citep{matias2014}.  An early (but still popular) approach in this direction is the stochastic blockmodel, which assumes that each agent possesses a latent, categorical class (or group membership). Nodes within each class are assumed to be stochastically equivalent in their connectivity behavior, meaning that the probability of two nodes to connect depends solely on their group memberships \citep{holland1983, sischka2022}. This family of models is attractive due to its simplicity in detecting and describing subgroups of nodes in networks. In many applications, however, discrete groupings fail to adequately represent the observed data, as agents behave more heterogeneously.
Moving from discrete to continuous latent variable network models, another prominent approach 
is the latent distance model. The latter postulates that agents are positioned in a latent Euclidean ``social space'', and that the closer they are within it, the more likely they are to form ties \citep{hoff2002}. More precisely, the classical latent distance model specifies the probability of observing an edge between nodes $i$ and $j$, conditional on $Z$, through
\begin{equation}
\mathbb{P}_\theta(Y_{ij} = 1 | \boldsymbol{Z}) = \frac{\exp\{\theta^\top x_{ij} - \lVert \boldsymbol{z}_{i} - \boldsymbol{z}_{j} \rVert \}}{1+\exp\{\theta^\top x_{ij}- \lVert \boldsymbol{z}_{i} - \boldsymbol{z}_{j} \rVert\}},
\end{equation}
where $\boldsymbol{Z} = (\boldsymbol{z}_1,...,\boldsymbol{z}_n)$ denotes the latent positions of the nodes in the $d$-dimensional latent space, and $\theta$ is the coefficient vector for the covariates $x_{ij}$. The latent positions $\boldsymbol{Z}$ are assumed to originate independently from a spherical Gaussian distribution, i.e.  $Z \sim N_d(0, \tau^2 \bm{I}_d))$, where $\bm{I}_d$ indicates a $d$-dimensional identity matrix. 

Latent distance models are particularly attractive for social networks in which triadic closure plays a major role, and where nodes with similar characteristics tend to form connections with each other (i.e.\ homophilic networks, see \citealp{rivera2010}). It is also possible to add nodal random effects to the model, to control for agent-specific heterogeneity in the propensity to form edges \citep{krivitsky2009}. The model then becomes
\begin{equation}
\mathbb{P}_\theta(Y_{ij} = 1 | \boldsymbol{Z}, a,b) = \frac{\exp\{\theta^\top x_{ij} - \lVert \boldsymbol{z}_{i} - \boldsymbol{z}_{j} \rVert + a_i + b_j \}}{1+\exp\{\theta^\top x_{ij}- \lVert \boldsymbol{z}_{i} - \boldsymbol{z}_{j} \rVert + a_i + b_j \}},
\end{equation}
 where $a = (a_1,...,a_n)$ and $b = (b_1,...,b_n)$ are node-specific sender and receiver effects that account for the individual agents' propensity to form ties, with $a \sim N_n({0},\tau^2_{a}\mathbf{I}_n)$ and $b \sim N_n({0},\tau^2_{b}\mathbf{I}_n)$. 

Despite its advantages and its fairly simple interpretation, a Euclidean latent space is unable to effectively approximate the behavior of networks where nodes that are similar in terms of connectivity behavior are not necessarily more likely to form ties \citep{hoff2008}, such as, e.g., many networks of amorous relationships \citep{ghani1997, bearman2004}.
More generally, the latent distance model tends to perform poorly for networks in which stochastic equivalence does not imply homophily and triadic closure, i.e., when nodes which behave similarly in terms of connectivity patterns towards the rest of the network do not necessarily have a higher probability of being connected among themselves. 
This is often the case in economics, where real-world networks can exhibit varying degrees and combinations of stochastic equivalence, triadic closure and homophily. Moreover, it is often \textsl{a priori} unclear which of these mechanisms are at play in a given observed network. In this context, agent-specific multiplicative random effects instead of the additive latent positions allow for simultaneously representing all these patterns \citep{hoff2005}. Further developments of this innovation have led to the modern specification of the Additive and Multiplicative Effects network model (AME,  \citealp{hoff2011}), which, from a matrix representation perspective, generalizes both the stochastic blockmodel and the latent distance model \citep{hoff2021}. 

\subsection{AME: Motivation and framework}
The AME approach can be motivated by considering that network data often exhibit first-, second-, and third-order dependencies.
\textsl{First-order effects} capture agent-specific heterogeneity in sending (or receiving) ties within a network. For example, in the case of companies and legal disputes, first-order effects can be viewed as the propensity of each firm to initiate (or be hit by) legal disputes. 
\textsl{Second-order effects}, i.e., reciprocity, describe the statistical dependency of the directed relationship between two agents in the network. In the previous example, this effect can be described as the correlation between (a) company $i$ initiating a legal dispute against company $j$ and (b) $j$ doing the same towards $i$. Of course, second-order effects can only occur in directed networks.
\textsl{Third-order effects} are described as the dependency within triads, defined as the connections between three agents, and relate to the triangular statistics previously illustrated in Figure \ref{fig:triangular_statistics}. How likely is it that ``a friend of a friend is also my friend''? Or, returning to the previous example: given that $i$ has legal disputes with $j$ and $k$, how likely are disputes to occur between $j$ and $k$?

The AME network model is designed to simultaneously capture these three orders of dependencies. More specifically, it extends the classical (generalized) linear modeling framework by incorporating extra terms into the systematic component to account for them. In the case of binary network data, we can make use of the Probit AME model. As is well known, the classical Probit regression model can be motivated through a latent variable representation in which $y_{ij}$ is the binary indicator that some latent normal random variable, say $L_{ij} \sim \mathcal{N}(\theta^\top\boldsymbol{x}_{ij} , \sigma^2)$, is greater than zero \citep{albert1993}. But an ordinary Probit regression model assumes that $L_{ij}$, and thus the binary indicators (edges) $y_{ij}$, are independent, which is generally inappropriate for network data. 
In contrast, the AME Probit model specifies the probability of a tie $y_{ij}$ from agent $i$ to agent $j$, conditional on a set of latent variables $W$, as
\begin{equation}
\label{eq:amen1}
\mathbb{P}(Y_{ij} = 1 | W) = 
\mathbf{\Phi}(\theta^\top\boldsymbol{x}_{ij} + e_{ij}), 
\end{equation}
where $\boldsymbol{\Phi}$ is the cumulative distribution function of the standard normal distribution, $\theta^\top \boldsymbol{x}_{ij}$ accommodates the inclusion of dyadic, sender, and receiver covariates, and $e_{ij}$ can be viewed as a structured residual, containing the latent terms in $W$ to account for the network dependencies described above. In the directed case, $e_{ij}$ is composed as
\begin{equation}
\label{eq:amen2}
e_{ij} = a_i + b_j + {u}_i{v}_j + \epsilon_{ij}.
\end{equation}
In this context, $a_i$ and $b_j$  are zero-mean additive effects for sender $i$ and receiver $j$ accounting for first-order dependencies, jointly specified as 
\begin{equation}
  \begin{split}
  (a_1,b_1),...,(a_n,b_n)\overset{\text{i.i.d.}}{\sim}  N_{2}(0,\Sigma_1),
  \end{split}
\quad\text{with}\quad
  \begin{split}
  \Sigma_1 =  \begin{pmatrix} \sigma_{a} & \sigma_{ab} \\ \sigma_{ab} & \sigma_{b} \end{pmatrix}
  \end{split}.
\end{equation}
The parameters $\sigma_a$ and $\sigma_b$ measure the variance of the additive sender and receiver effects, respectively, while $\sigma_{ab}$ relates to the covariance between sender and receiver effects for the same node. 
Going back to (\ref{eq:amen2}), $\epsilon_{ij}$ is a zero-mean residual term which accounts for second order dependencies, i.e.\ reciprocity. More specifically, it holds that
\begin{equation}
  \begin{split}
\{(\epsilon_{ij},\epsilon_{ji}):i<j\} \overset{\text{i.i.d.}}{\sim} N_2(0,\Sigma_2),
  \end{split}
\quad\text{with}\quad
  \begin{split}
  \Sigma_2 = \sigma^2
\begin{pmatrix} 1 & \rho \\ \rho & 1 \end{pmatrix}
  \end{split},
\end{equation}
where $\sigma^2$ denotes the error variance and $\rho$ determines the correlation between $\epsilon_{ij}$ and $\epsilon_{ji}$, thus quantifying the tendency towards reciprocity.
Finally, $\mathbf{u}_i$ and $\mathbf{v}_j$ in (\ref{eq:amen2}) are $d$-dimensional multiplicative sender and receiver effect vectors that account for third-order dependencies, and for which $({u}_1,{v}_1),...,({u}_n,{v}_n) \sim \mathcal{N}_{2d}({0},\Sigma_3)$ holds.


As noted above, AME is able to represent a wide variety of network structures, generalizing several other latent variable model classes. This generality comes at the price of a high level of complexity for the estimated latent structure. This can make the model class a sub-optimal choice if one wants to interpret the latent structure with respect to, e.g., clustering. On the other hand, its flexibility makes it an ideal fit when the underlying network dependencies are unknown, and the researchers’ interest mainly lies in evaluating and interpreting the effect of dyadic and nodal covariates on tie formation while controlling for network effects. This strength has led to AME being used for several applications of this type \citep{koster2018,minhas2019,Minhas_Dorff_Gallop_Foster_Liu_Tellez_Ward_2021,Dorff_Gallop_Minhas_2020}. We next showcase the AME framework by applying it to the world foreign exchange activity network as of 1900, originally introduced and studied by \cite{flandreau2005,flandreau2009}. This application highlights how using AME instead of classical regression can allow us to reconsider existing, influential answers to relevant questions via replication. 

\subsection{Application to the global foreign exchange activity network} 
\begin{figure}[t]

{\centering \includegraphics[width=0.65\linewidth, trim=0.6cm 7.5cm 0.6cm 7.5cm, clip]{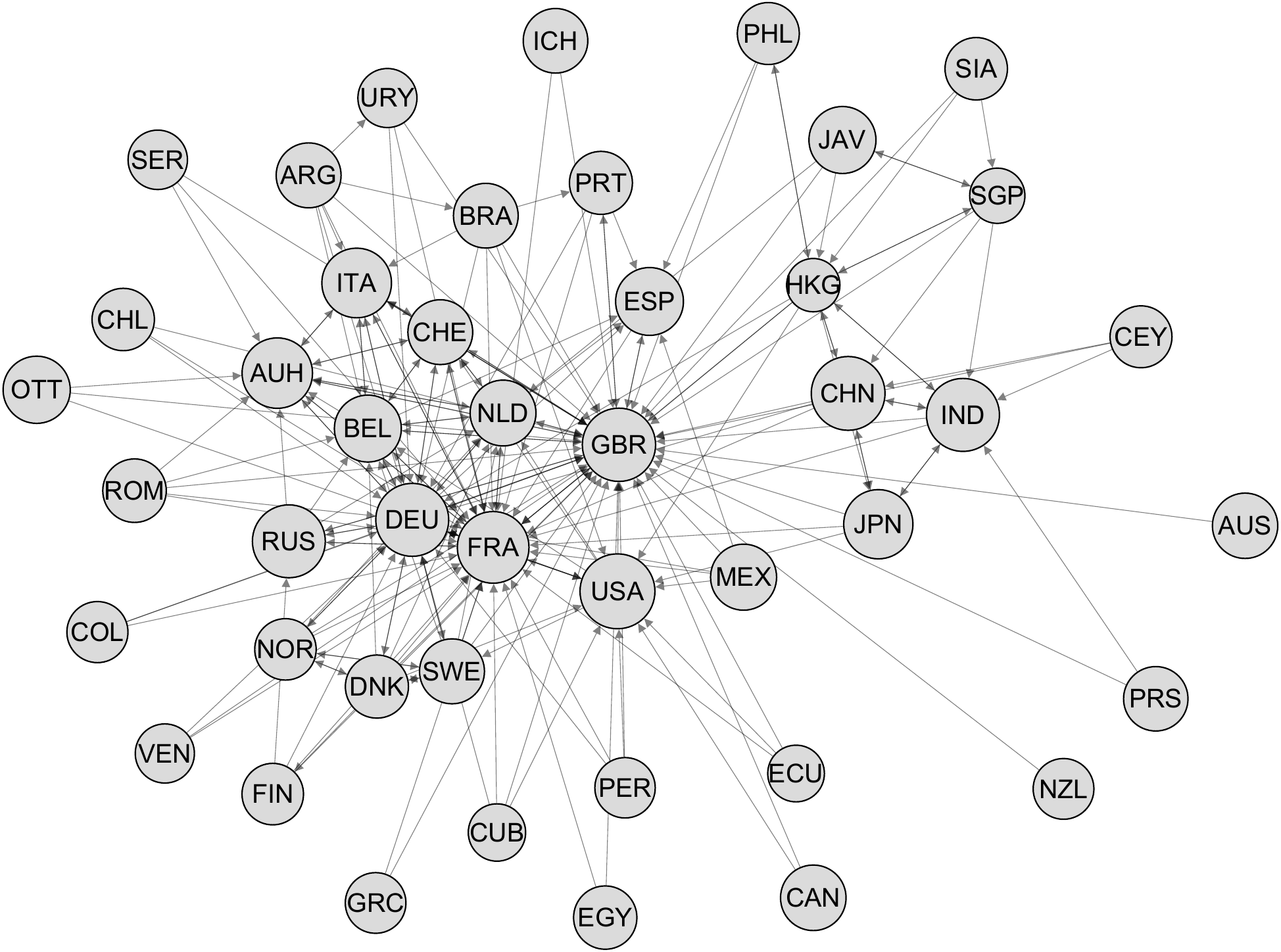} 

}
\caption{Illustration of the global foreign exchange activity network in 1900. Countries are labeled by their ISO 3166-1 codes, and an edge from node $i$ to node $j$ indicates active trading of the currency from country $j$ within a financial center of country $i$.
}\label{fig:forex}
\end{figure}

In 1900, every financial center featured a foreign exchange market were bankers bought and sold foreign currency against the domestic one. Foreign exchange market activity was monitored in local bulletins, which allowed \cite{flandreau2005} to collect a global dataset with all currencies used in the world at that time. In the resulting network structure, laid out in Figure \ref{fig:forex},  countries are nodes, and a (directed) edge from country $i$ to country $j$ occurs if the currency of country $j$ was actively traded in at least one financial center within country $i$. From the graph representation, laid out using a variant of the Yifan Hu force-directed graph drawing algorithm \citep{hu2005}, we observe that the most actively traded currencies at the time belonged to large European economies, such as Great Britain, France and Germany. To determine the drivers of currency adoption, \citet{flandreau2009} model this network as a function of several covariates by employing ordinary binary regression. As we show, it is possible to use AME to pursue the same goal while taking network dependencies into account. 

We specify the AME model as in (\ref{eq:amen1}), using directed edges $y_{ij}$ as response variable.
The nodal covariates we use, sourced from and described in detail in the replication materials of \cite{flandreau2009}, are (log)per-capita GDP, democracy index score, coverage of foreign currencies traded in the country, and an indicator of whether the country's currency was on the gold standard. We also include, as dyadic covariates, the distance between two countries as well as their total trade volume. As specified in (\ref{eq:amen2}), the structured residual term $e_{ij}$ comprises additive effects $a_i$ and $b_j$ for each node, which capture country-specific propensities to send and receive ties, respectively. Multiplicative effects $\boldsymbol{u}_i$ and $\boldsymbol{v}_j$ are included to account for third order dependencies. We here set the dimensionality of the multiplicative effects to two, which we assume to be sufficient given the relatively small size of the network. 

To estimate the AME model, we make use of the \texttt{R} package \texttt{amen} \citep{hoff2015}. As the likelihood involves intractable integrals arising from the combination of the transformation and dependencies induced by the the model, closed form solutions are not available. The package thus uses reasonably standard Gibbs sampling algorithms to provide Bayesian inference on the model parameters. More details on the estimation routine can be found in \cite{hoff2021}.

The results of the analysis, as well as, for comparison's sake, a Probit regression including the same covariates but ignoring network dependencies,  are displayed in Table \ref{table:amen}. Note that the classical Probit regression model can be seen as a special case of AME Probit in which both additive and multiplicative node-specific effects are omitted. Additional model diagnostics and goodness of fit measures, together with the estimated variance and covariance parameters, are provided in the Supplementary Material. The estimated coefficients (for both models) can be interpreted as in standard Probit regression: For the nodal covariate {per-capita GDP},  for example, a unit increase in the log-per-capita GDP for country $i$ corresponds to a decrease of 0.453 in the linear predictor, therefore negatively influencing the expected probability of the country to \textit{send} a tie. 
The same unit increase in the log-per-capita-gdp for country $i$ corresponds to an increase of 0.426 in the linear predictor, and has therefore a positive impact on the expected probability of that country to \textit{receive} a tie. 
In the case of a dyadic covariate, such as {distance}, a unit increase in distance between two countries leads to a decrease of 1.019 in the linear predictor, resulting in a decrease in the expected probability of the two countries to form a tie in either direction. 
Overall, we find that the principal drivers of the formation of a tie between $i$ and $j$ are the magnitude of the foreign exchange coverage of the two countries involved, the distance between them, and their reciprocal trade volume. These results correspond to the thesis of \cite{kindleberger1967} and to \cite{flandreau2009}, who suggest that the most important determinants of international adoption for a currency are size and convenience of use. At the same time, we note that, as for the ERGM in the arms trade example, the results of the Probit and AME model differ in several regards. In particular, several effects are statistically significant in the Probit but not significant in the AME model. Indeed, unacknowledged network dependence can cause downward bias in the estimation of standard errors, leading to spurious associations \citep{lee_network_2021}. This finding once again highlights how accounting for network dependencies can make a difference when it comes to the substantive results. 

As a final note, we add that in this case we went with AME over ERGM as our interest lies in answering the research questions addressed by \cite{flandreau2005,flandreau2009}, that is assessing the effect of the exogenous covariates in Table \ref{table:amen} on tie formation. AME allows us to do that without specifying the configuration of the endogenous network mechanisms at play, which are instead accounted for through the imposed latent structure. If, on the other hand, the researcher expects some specific network effects to play a role, and wishes to test for their presence and measure their influence on network formation, the ERGM may be a better tool. The latter model class can, for example, directly answer questions such as ``Does the fact that both countries A and B trade the currency of country C influence the probability of A and B to be connected? And if so, to what extent?''. AME, on the other hand, is limited to accounting for those effects via the latent variables, without explicitly identifying them, to provide unbiased inference for the covariate effects. The choice between the two model classes is thus a matter of what assumptions can be made about the network and where the researcher's interest lies.

\begin{table}
\begin{center}
\begin{tabular}{rl c c}
\hline
 & & {AME} & Classical Probit \\
\hline
& Intercept    & $-4.845 \; (5.310)$       & $-3.211 \; (1.580)^{*}$   \\
\ldelim\{{4}{1.3cm}[Sender] & Gold standard     & $-0.629 \; (0.397)$       & $-0.354 \; (0.155)^{*}$   \\
& log-GDP per-capita  & $-0.453 \; (0.419)$       & $-0.259 \; (0.152)$       \\
& Democracy index  & $-0.033 \; (0.064)$       & $-0.025 \; (0.026)$       \\
& Currency coverage & $1.418 \; (0.405)^{***}$  & $0.470 \; (0.137)^{***}$  \\
\ldelim\{{4}{1.6cm}[Receiver] & Gold standard    & $-0.599 \; (0.667)$       & $-0.468 \; (0.191)^{*}$   \\
& log-GDP per-capita  & $0.426 \; (0.703)$        & $0.240 \; (0.159)$        \\
 & Democracy index  & $0.121 \; (0.102)$        & $0.066 \; (0.019)^{***}$  \\
& Currency coverage & $2.734 \; (0.691)^{***}$  & $1.363 \; (0.181)^{***}$  \\
\ldelim\{{2}{1.35cm}[Dyadic] & Distance    & $-1.019 \; (0.151)^{***}$ & $-0.471 \; (0.064)^{***}$ \\
& log-trade volume & $0.488 \; (0.081)^{***}$  & $0.346 \; (0.036)^{***}$  \vspace{0.1cm}\\
\hline
\multicolumn{3}{l}{\scriptsize{$^{***}p<0.001$; $^{**}p<0.01$; $^{*}p<0.05$}}
\end{tabular}
\caption{Estimated coefficients and related standard errors (in parentheses) for the AME model and the corresponding Probit model for the global foreign exchange activity network in 1900.}
\label{table:amen}
\end{center}
\end{table}


\section{Conclusion}
\label{sec:discussion}

Complex dependencies are ubiquitous in the economic sciences \citep{chiarella2005,flaschel2008}, and many economic interactions can be naturally perceived as networks. This area of research has thus received considerable interest in recent years. However, this attention has not yet been accompanied by a corresponding general take-up of empirical research methods tailored towards networks. Instead, researchers either develop their own estimators to reproduce the features of their theoretical network models, or use standard regression methods that assume conditional independence of the edges in the network. Against this background, this paper seeks to provide a hands-on introduction to two statistical models which account for network dependencies, namely the Exponential Random Graph Model (ERGM) and the Additive and Multiplicative Effects network model (AME). These two classes serve different purposes: While the ERGM is most appropriate when explicitly interested in testing the effects of endogenous network structure, the AME model allows one to control for network dependencies while substantively focusing on estimating the effects of exogenous covariates of interest. We present the statistical foundations of both models, and demonstrate their applicability to economic networks through examples in the international arms trade and foreign currency exchange, showing that modeling network dependencies can alter the substantive results of the analysis. We, moreover, provide the full data and code necessary to replicate these exemplary applications. We explicitly encourage readers to use these replication materials to get started with analyzing economic networks via ERGM and AME, beginning with the examples covered here to then transfer the code and methods to their own research.

We especially want to encourage the use of such methods as not accounting for interdependence between observations when it exists can lead to biased estimates and spurious findings. Our two applications demonstrate that this bias can result in very different empirical results, and thus affect substantive conclusions. It is therefore vital to account for network structure when studying interactions between economic agents such as individuals, firms, or countries, regardless of whether one is substantively interested in this structure. 
As shown by \citet{lee_network_2021}, our applications are just two examples of how unaccounted dependence in the observed data may lead to spurious findings.

At the same time, this paper can only serve as an introduction to statistical network data analysis in economics. We covered two general frameworks in this realm, but, in the interest of brevity, focused only on their simplest versions that apply to networks observed at only one time point and with binary edges. However, both frameworks have been extended to cover more general settings. For the ERGM, there are extensions for longitudinal data \citep{Hanneke_Fu_Xing_2010}, distinguishing between edge formation and continuation \citep{krivitsky2014separable}, as well as to settings where edges are not binary but instead count-valued or signed \citep{krivitsky2012exponential,fritz2022exponential}. As for AME, approaches for longitudinal networks are described by \cite{minhas2016}, while versions for undirected networks as well as for non-binary network data are presented by \cite{hoff2021}. Both the ERGM and the AME frameworks are thus flexible enough to cover a wide array of potential economic interactions. We believe that increasingly adopting these methods will, in turn, aid our understanding of these interactions. 

\textbf{}

\newpage
\bibliographystyle{chicago}
\bibliography{refs}


\newpage

\renewcommand{\thefootnote}{\fnsymbol{footnote}} 

\begin{center}
	{\Large{{Supplementary Material}} \\
 \medskip
 \Large{\textbf{Dependence matters: Statistical models to identify the drivers of tie formation in economic networks}}}  \\ 
	Giacomo De Nicola$^{1,}$\footnote[1]{Corresponding author:  giacomo.denicola@stat.uni-muenchen.de}, Cornelius Fritz$^2$, Marius Mehrl$^3$, Göran Kauermann$^{1}$\hspace{.2cm}\\
	\vspace{0.3cm}
	Department of Statistics, LMU Munich$^1$ \\
 Department of Statistics, Pennsylvania State University$^2$ \\
School of Politics and International Studies, University of Leeds$^3$\hspace{.2cm}
\end{center}

\setcounter{section}{0}
\setcounter{page}{1}

\makeatletter 
\setcounter{figure}{0}
\setcounter{table}{0}
\renewcommand{\thefigure}{S.\@arabic\c@figure}
\renewcommand{\thetable}{S.\@arabic\c@table}
\renewcommand{\thesection}{S.\@arabic\c@section}

\makeatother

\setcounter{equation}{0}

The following contains technical details and supplementary information to the manuscript. The full data and code to reproduce the analysis and to aid the reader in applying the presented models to their own data can be found in our {Github repository}, available at \href{https://github.com/gdenicola/statistical-network-analysis-in-economics}{\texttt{https://github.com/gdenicola/statistical-network-analysis-in-economics}}.

\section{Details on the $p_1$ model}

To represent reciprocity, the $p_1$ model assumes dyads, defined by $(Y_{ij}, Y_{ji})$, to be independent of one another. The resulting bivariate distribution of each dyad $(Y_{ij}, Y_{ji})$ comprises three parameters and a constraint: 
\begin{align*}
    m_{ij} &= \mathbb{P}(Y_{ij} = Y_{ji} = 1)\\
    a_{ij} &= \mathbb{P}(Y_{ij} = 1, Y_{ji} = 0)\\
    e_{ij} &= \mathbb{P}(Y_{ij} = Y_{ji} = 0) \\
   m_{ij} &+  a_{ij} +  a_{ji} + e_{ij}  = 1.
\end{align*}
The joint distribution of the network is then given by: 
\begin{align}
\label{eq:p_1}
    \mathbb{P}_{\theta}(\bm{Y} = \bm{y}) &= \prod_{i<j} m_{ij}^{y_{ij}y_{ji}}  a_{ij}^{y_{ij}(1-y_{ji})}  a_{ji}^{(1-y_{ij})y_{ji}}  e_{ij}^{(1-y_{ij})(1-y_{ji})} \nonumber\\
        &= \frac{\exp\left\{ \sum_{i<j} \rho_{ij} y_{ij}y_{ji}  + \sum_{i \neq j} \theta_{ij} y_{ij} \right\}}{\kappa(\theta)} , 
\end{align}
where $\rho_{ij} = \log\left(m_{ij} e_{ij} / a_{ij}^2\right)$ for $i<j$, $\theta_{ij} =  \log\left( a_{ij}/n_{ij}\right)$ for $i \neq j$, and $\kappa(\theta)$ is the normalizing constant. To estimate the parameters in \eqref{eq:p_1} we need to introduce some homogeneity assumption to avoid overparametrization. Following \citetsupp{holland1981}, we assume 
\begin{align*}
    \rho_{ij} &= \theta_{\text{Rep}} ~\forall ~ i<j \\
    \theta_{ij} = \theta_{\text{Edges}}& + \theta_{\text{Out},i}+ \theta_{\text{In},j} ~\forall ~ i\neq j \\
   \sum_{i = 1}^n \theta_{\text{Out},i} &=    \sum_{i = 1}^n \theta_{\text{In},i} = 0,
\end{align*}
where $\theta_{\text{Rep}}$ is the global effect of reciprocity, $\theta_{\text{Edges}}$ quantifies the general sparsity in the network, and $\theta_{\text{Out},i}$ and $\theta_{\text{In},i}$ indicate the general tendency for all agents $i \in \{1, ..., n\}$ to form out- or in-going ties. Combining these homogeneity assumptions with \eqref{eq:p_1} yields 
\begin{align}
    \mathbb{P}_{\theta}(\bm{Y} = \bm{y}) \propto \exp\Big\{ &\theta_{\text{Rep}} s_{\text{Rep}}(\bm{y})  +  \theta_{\text{Edges}} s_{\text{Edges}}(\bm{y})  + \nonumber\\&\sum_{i = 1}^n \theta_{\text{Out,i}} s_{\text{Out},i}(\bm{y}) + \sum_{i = 1}^n \theta_{\text{In,i}} s_{\text{In},i}(\bm{y})   \Big\}, \label{eq:p_1_hom}
\end{align}
where $s_{\text{Rep}}(\bm{y}) = \sum_{i<j} y_{ij}y_{ji}$ is the number of reciprocal ties, $s_{\text{Edges}}(\bm{y}) =  \sum_{i = 1}^n  \sum_{j \neq i} y_{ij}$ the number of edges, and $s_{\text{Out},i}(\bm{y})$ and $s_{\text{In},i}(\bm{y})$ the number of out- and in-going ties of agent $i$. 

Note that all these statistics are functions of the observed network and are the sufficient statistics, i.e. they contain all necessary information for determining all coefficients in \eqref{eq:p_1_hom}. This sufficiency principle translates to $s_{\text{Out},i}(\bm{y}) = s_{\text{Out},j}(\bm{y}) \Rightarrow \theta_{\text{Out},i} = \theta_{\text{Out},j}$. This, in turn, allows us to write all agent-specific terms in \eqref{eq:p_1_hom} in terms of degree statistics: 
\begin{align}
    \mathbb{P}_{\theta}(\bm{Y} = \bm{y}) \propto \exp\Big\{ &\theta_{\text{Rep}} s_{\text{Rep}}(\bm{y})  +  \theta_{\text{Edges}} s_{\text{Edges}}(\bm{y})  + \nonumber\\&\sum_{i = 1}^n \theta_{\text{Outdeg,i}} s_{\text{Outdeg},i}(\bm{y}) + \sum_{i = 1}^n \theta_{\text{Indeg,i}} s_{\text{Indeg},i}(\bm{y})   \Big\}, \label{eq:p_1_hom_deg}
\end{align}
where $s_{\text{Outdeg},i}(\bm{y})$ and $s_{\text{Indeg},i}(\bm{y})$ are statistics counting the actors with out/in-degree $i$ in $\bm{y}$.  

\section{Further results of the application to the international arms trade network }

\subsection{Model diagnostics and goodness of fit}
\label{sec:annex_ergm}

\begin{figure}[]
\centering 
         \includegraphics[page=2,width=0.45\linewidth]{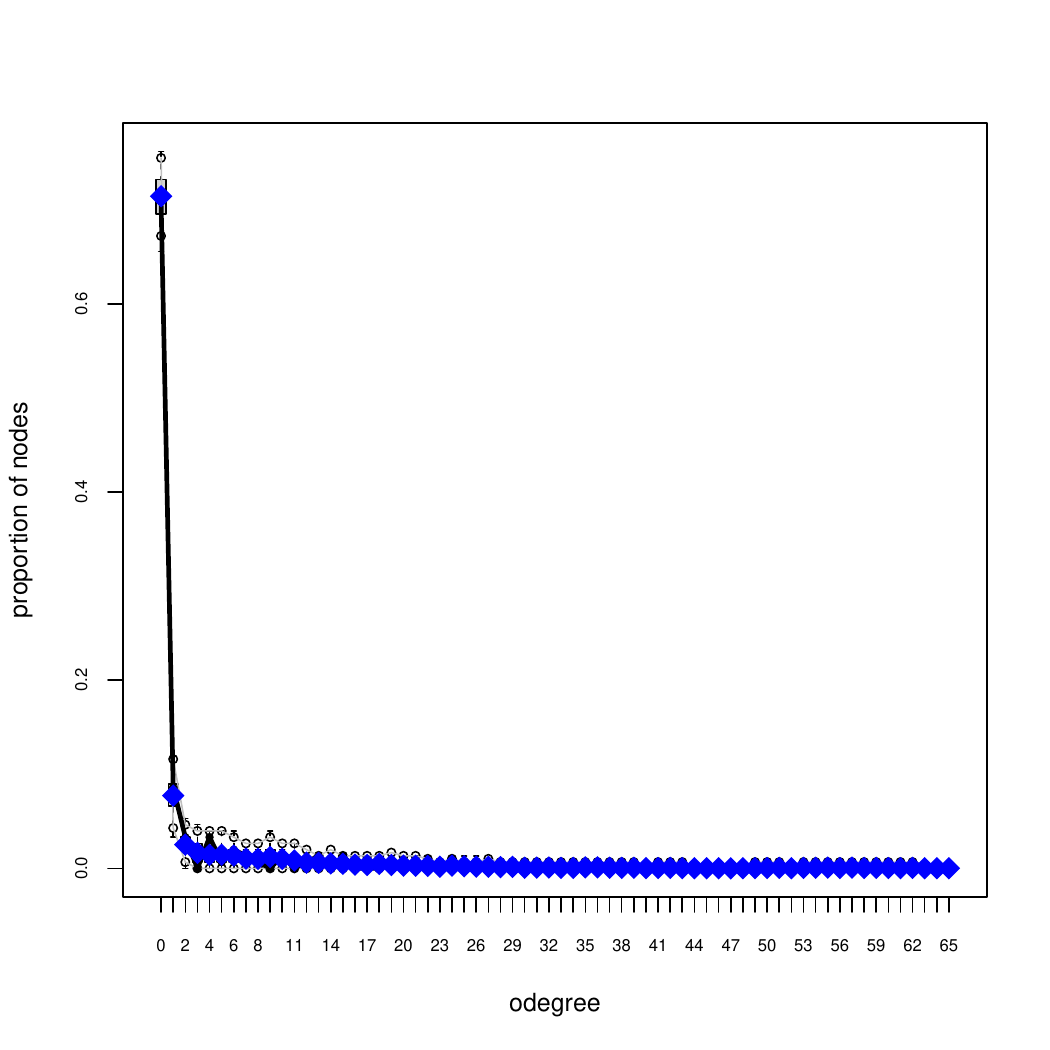}
         \includegraphics[page=3,width=0.45\linewidth]{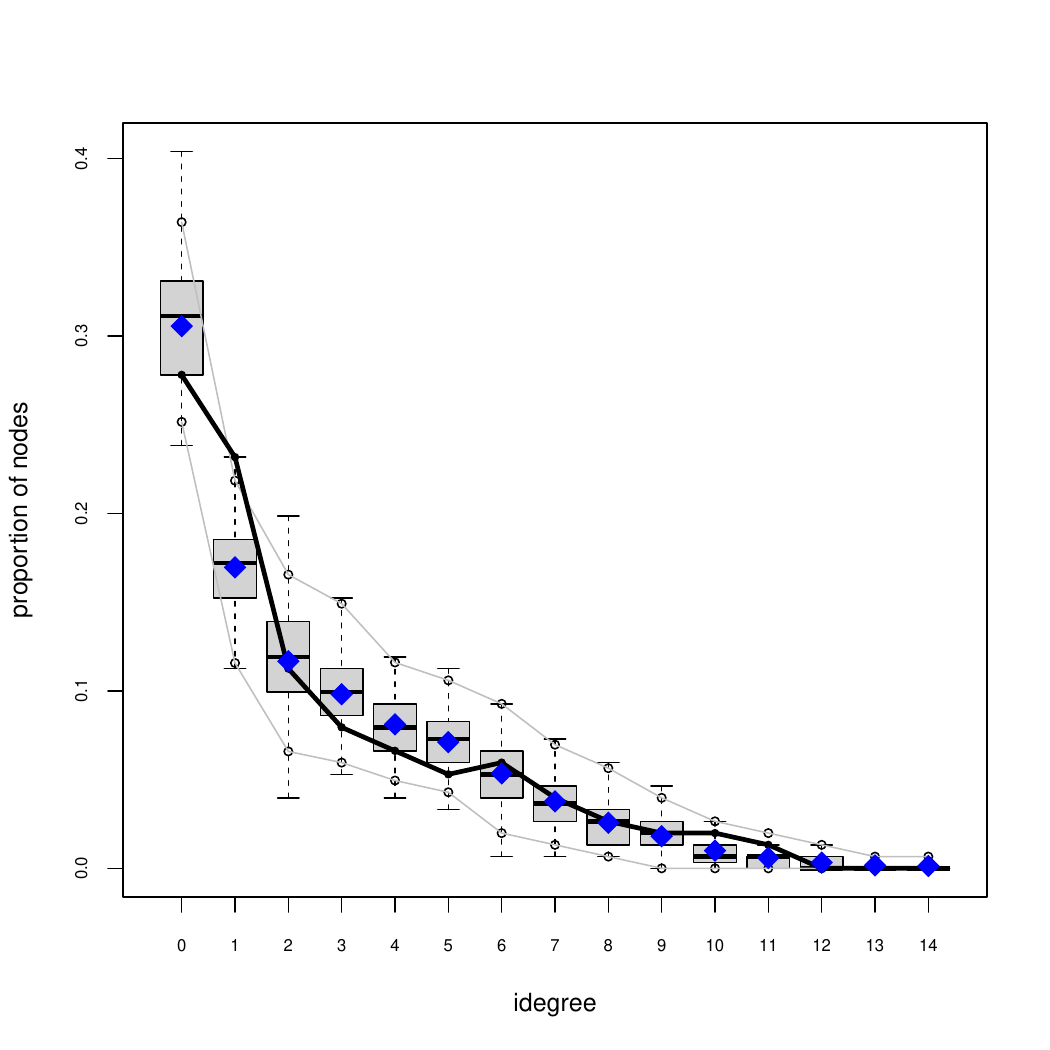}
         \includegraphics[page=4,width=0.45\linewidth]{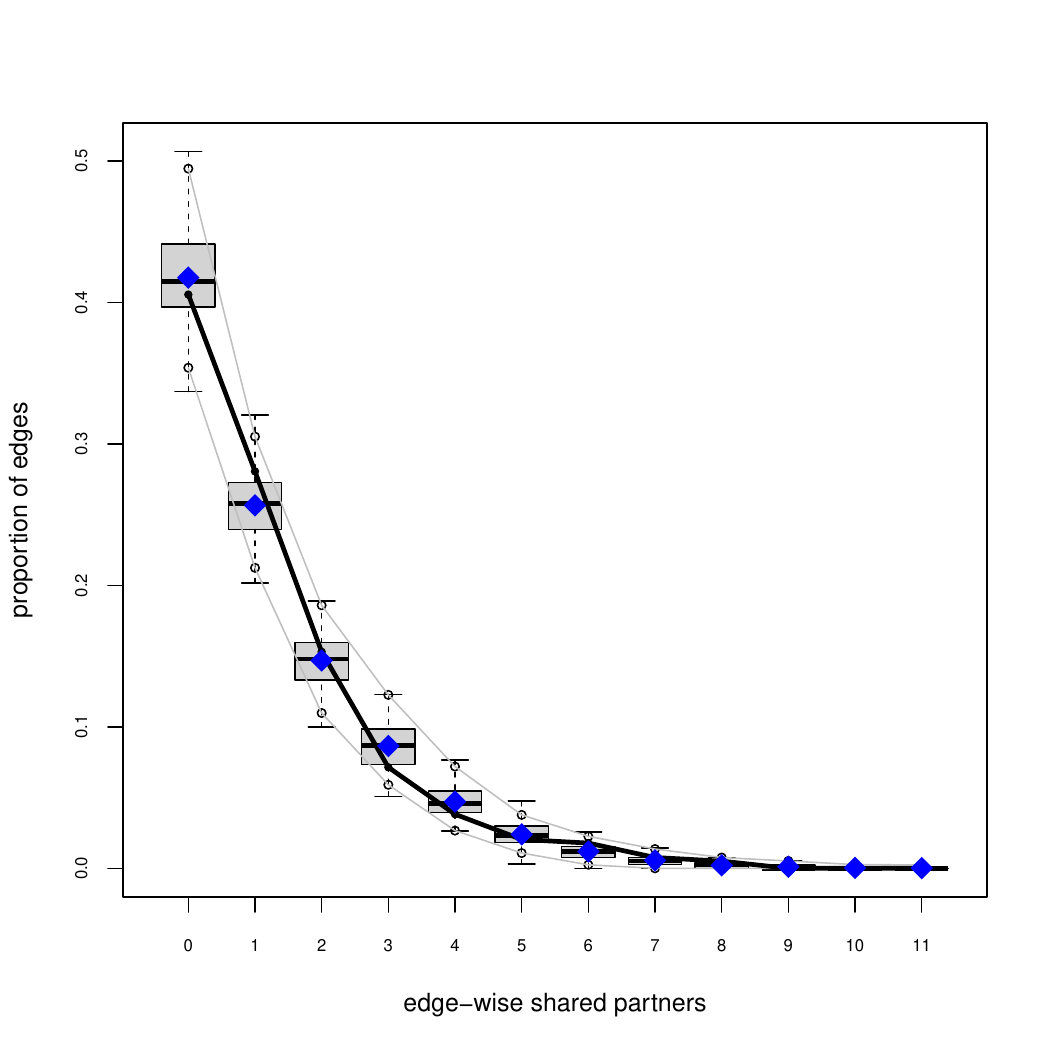}
         \includegraphics[page=5,width=0.45\linewidth]{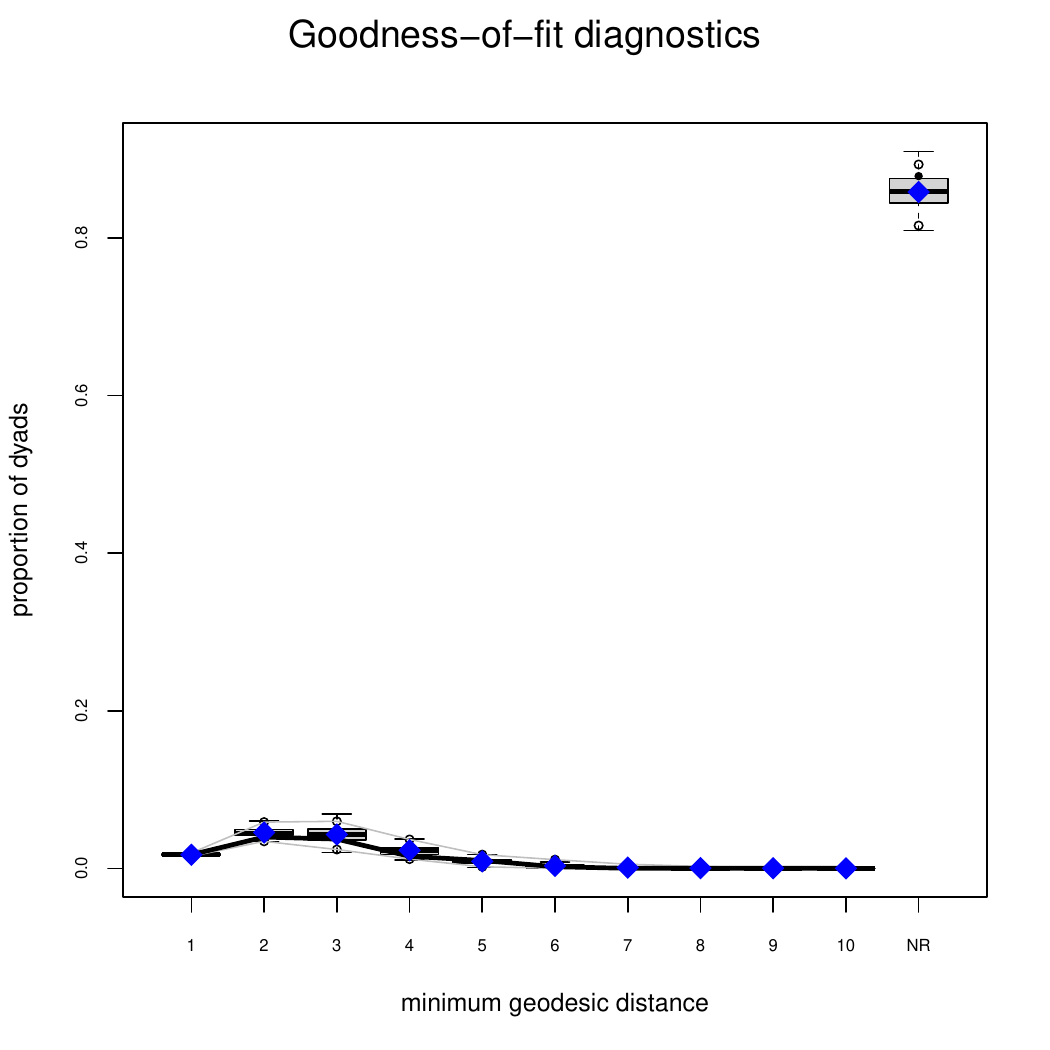}
\caption{Goodness of fit plots for the application to the international arms trade network in 2018. The metrics show a reasonably good fit for our model.}\label{fig:ergm_gof}
\end{figure}

To evaluate the goodness of fit of of an estimated ERGM, the standard approach is to compare the statistics observed in the real world network with the distribution of the same statistics calculated on networks simulated from the model \citepsupp{hunter_goodreau_handcock_2008}.
The results of this comparison are depicted in Figure \ref{fig:ergm_gof}, where we investigate the goodness of fit of the ERGM estimated on the 2018 international arms trade network.      
The black line in each subfigure represents the distribution of the respective network statistic observed in the real network. For instance, the top right figure indicates that approximately 10\% of all nodes (countries) in the network had an in-degree of 2. The boxplots then show the distribution of each value of a given statistic over the simulated networks. A good model will thus generally result in boxplots that include the observed values of the network statistics under consideration. In Figure \ref{fig:ergm_gof}, this is almost always the case, though it is also visible that the real network included less countries with an in-degree of 0 but more with an in-degree of 1 than the large majority of networks simulated from the ERGM.

\begin{figure}[t!]
\centering 
\includegraphics[width=0.45\linewidth]{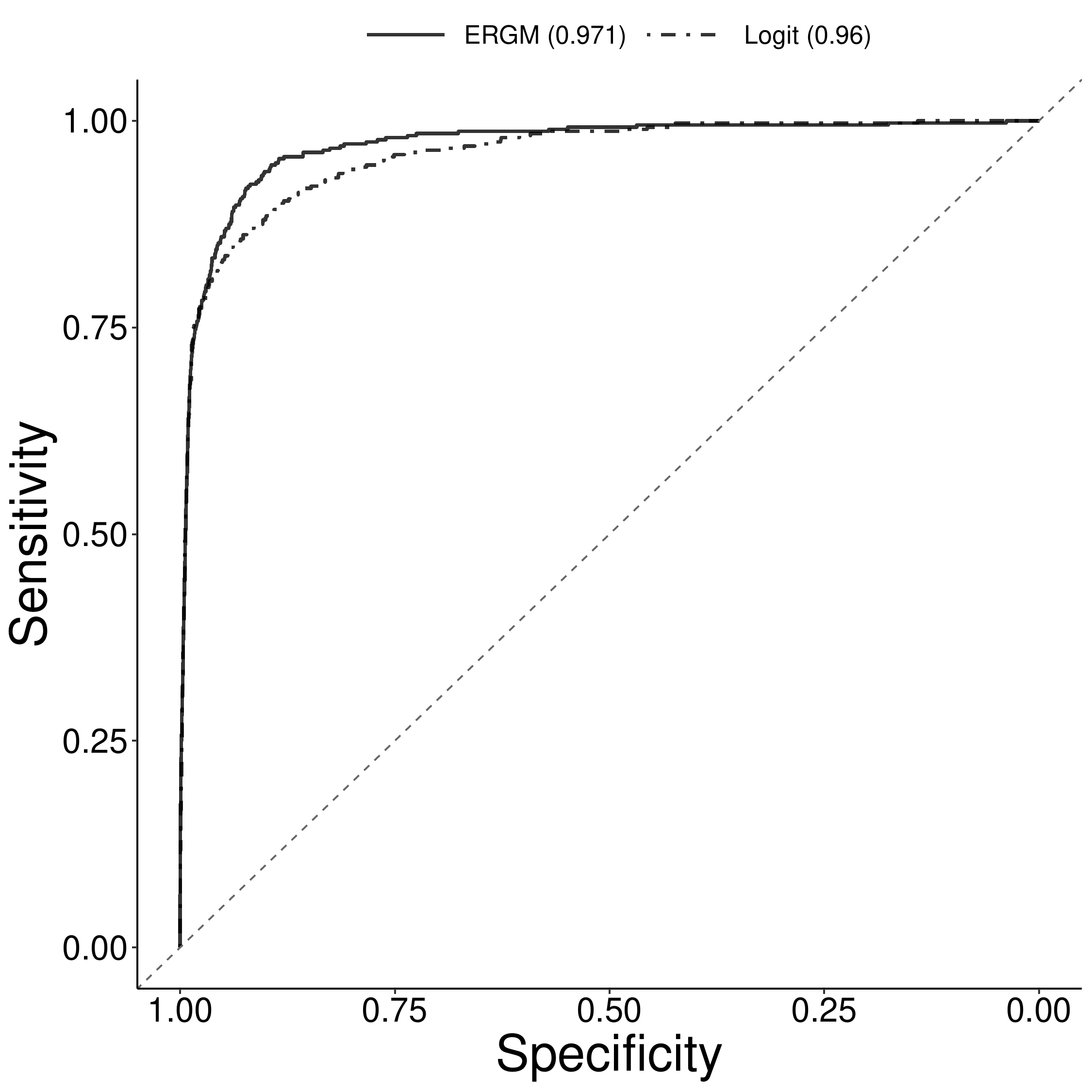}
\includegraphics[width=0.45\linewidth]{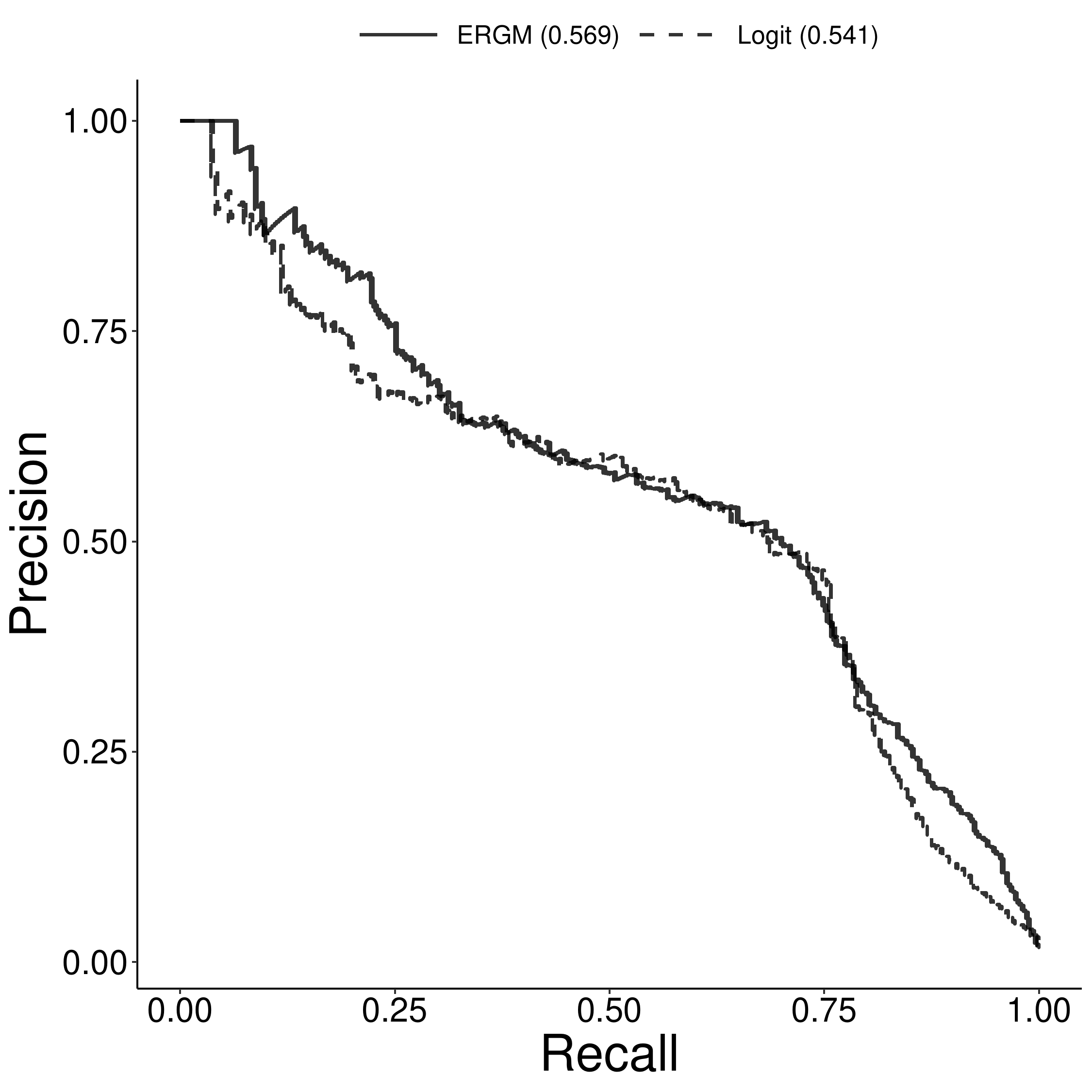}
\caption{In-sample prediction performance of the two models in Table 1. Left panel: ROC. Right panel: PR curve. ERGM outperforms Logit in both metrics.}\label{fig:ergm_mcmc}
\end{figure}

In Figure \ref{fig:ergm_mcmc}, we further compare the performance of the fully specified ERGM against that of the logistic regression model including the same exogenous covariates as the ERGM, but of course omitting all endogenous network statistics. We already noted that the ERGM appears to do a better job at in-sample prediction than the Logit model given its lower AIC and BIC values. Figure \ref{fig:ergm_mcmc} documents both models' respective areas under the receiver-operator (ROC) and precision-recall curves (PR). Here, a higher value of each curve indicates better predictive performance, and again, the ERGM appears to outperform the logistic regression model for both metrics. Figure \ref{fig:ergm_mcmc} thus offers further evidence that in the case of the international arms trade, model performance is improved by accounting for endogenous network effects.    

\subsection{Robustness checks}
\label{sec:robustness_check}

\begin{figure}[t!]
\centering 
\begin{subfigure}[c]{\textwidth}
   \includegraphics[width=0.45\linewidth]{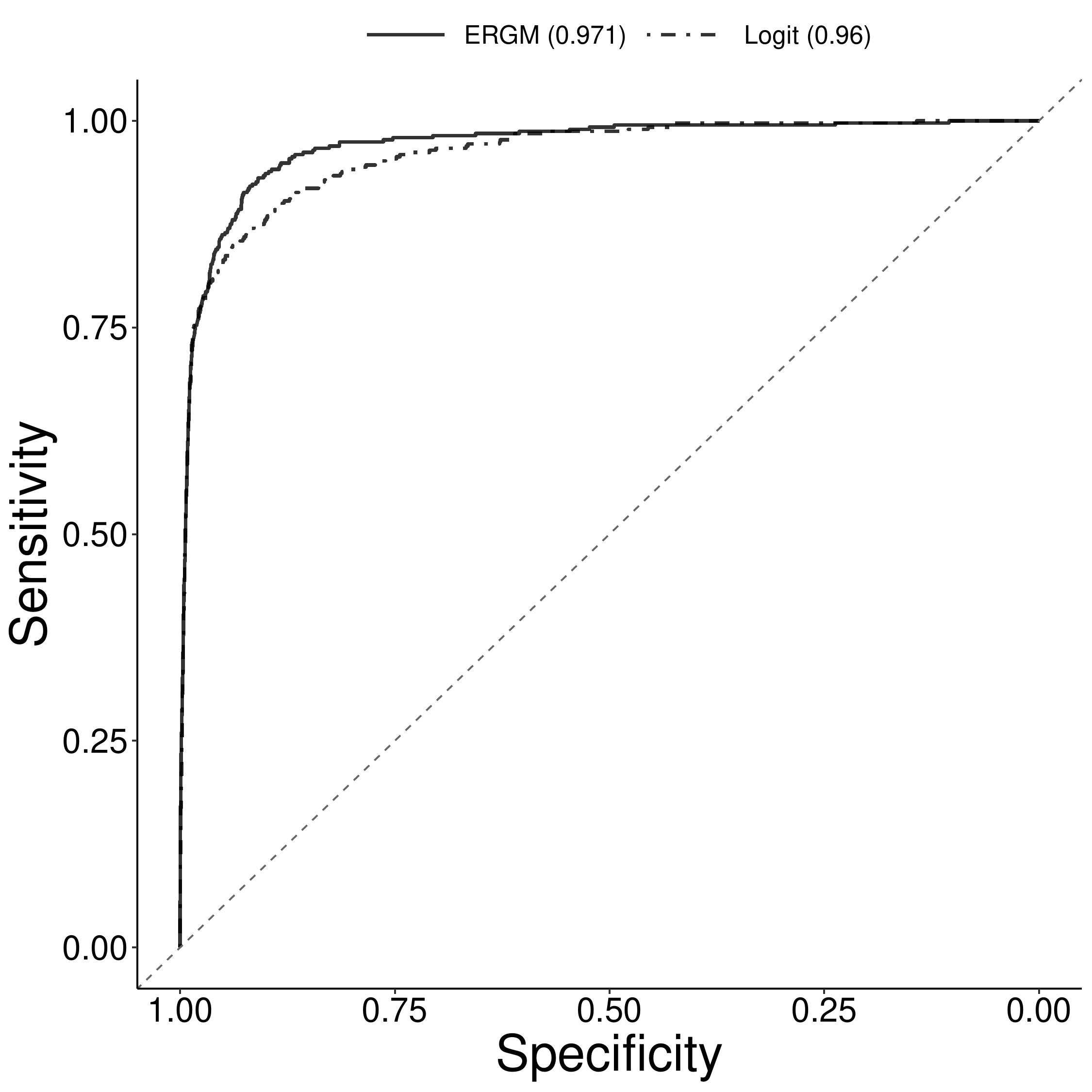}
\includegraphics[width=0.45\linewidth]{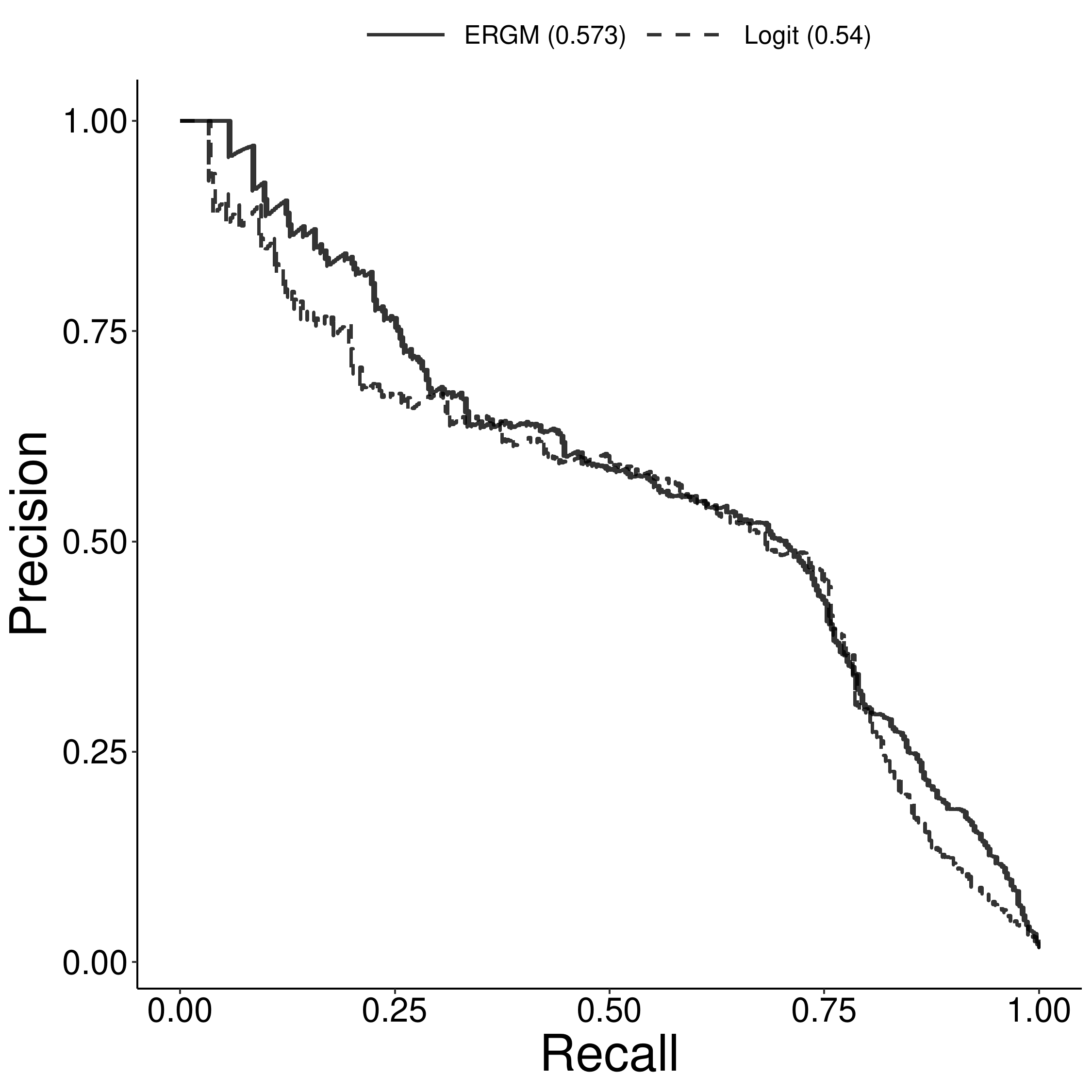}
	\subcaption{s = 4}
\end{subfigure}
\begin{subfigure}[c]{\textwidth}
   \includegraphics[width=0.45\linewidth]{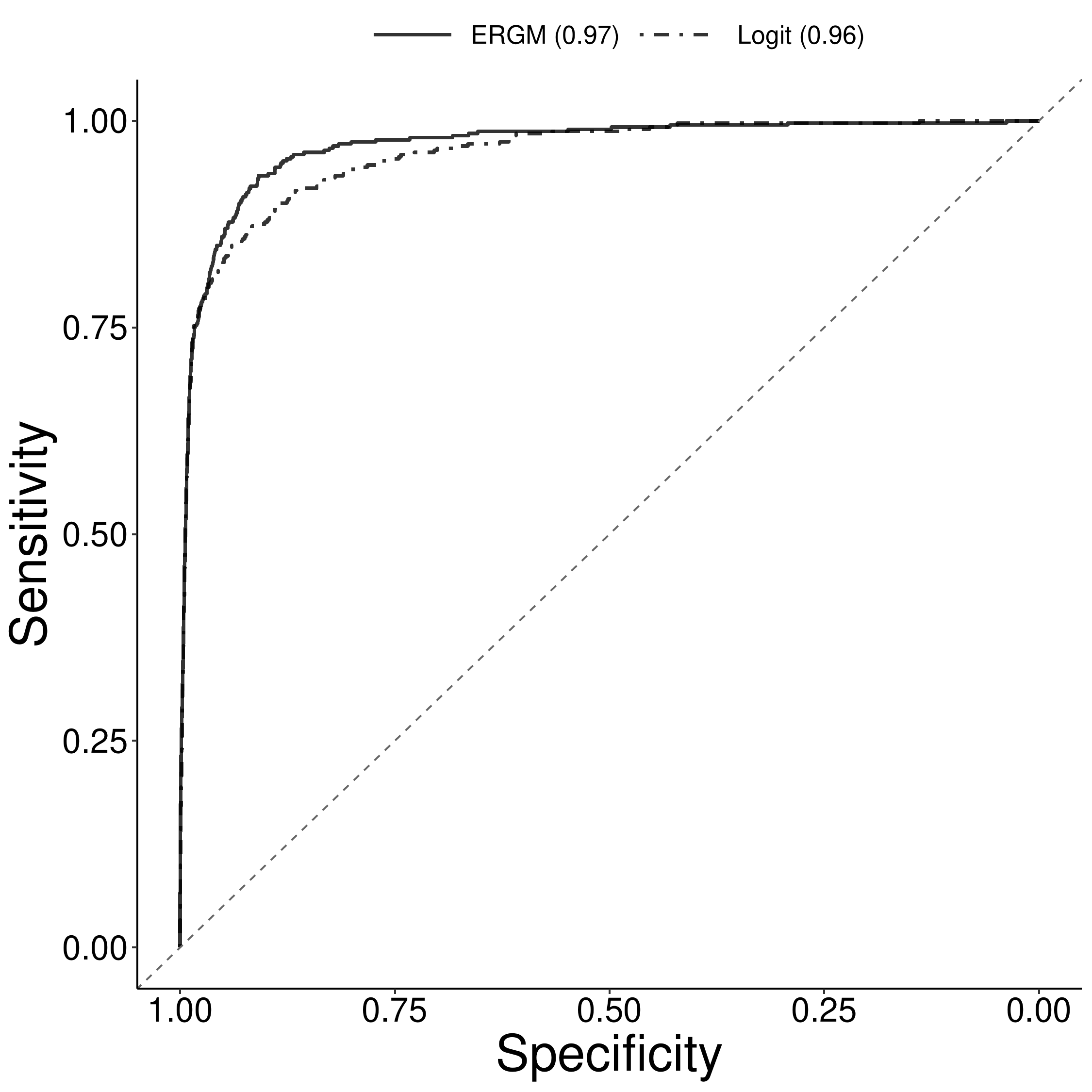}
\includegraphics[width=0.45\linewidth]{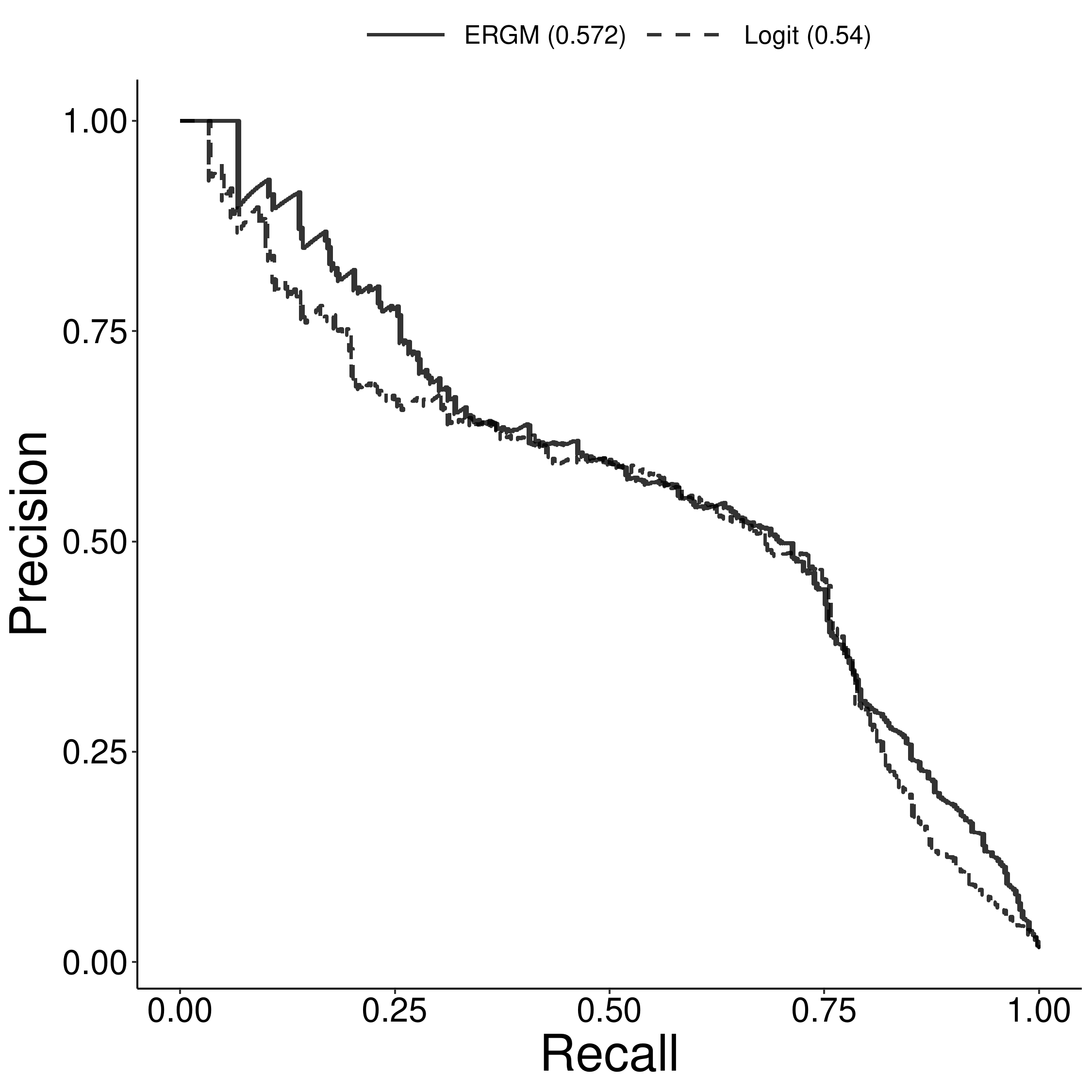}
	\subcaption{s = 5}
\end{subfigure}
\caption{In-sample prediction performance of the four models in Table \ref{table:coefficients_sm}. Upper row: Results when lagging the covariates four years. Lower row: Results when lagging the covariates five years. Left column: ROC. Right column: PR curve. ERGM outperforms Logit in both cases and for both metrics.}\label{fig:ergm_mcmc_sm}
\end{figure}

In the analysis presented in Section 3.3, we lag all covariate information by three years. Since this choice is based on the heuristic that there are considerable delays between the order and delivery date of arms, we next show the results of the ERGM with different time lags, namely four and five years, in Table \ref{table:coefficients_sm}. From these results, we conclude that the results are robust to different specifications of the time lags between the modeled network and the exogenous covariates. Moreover, one can observe that the ERGM consistently achieves lower AIC and BIC values compared to its classical logistic counterpart, underpinning the finding that accounting for network dependencies is necessary for analyzing the international arms trade. We further corroborate this point through a ROC as well as a PR analysis, both shown in Figures \ref{fig:ergm_mcmc_sm}(a) and (b). 

\begin{table}
\begin{center}
\begin{tabular}{l c c c c}
\hline
 & ERGM & Logit & ERGM & Logit \\
 Covariate Lag& 4 years & 4  years & 5 years & 5 years \\
\hline
Edges              & $-15.46 \; (2.10)$ & $-28.22 \; (1.73)$ & $-13.82 \; (3.19)$ & $-23.11 \; (3.04)$ \\
Repetition         & $3.25 \; (0.14)$   & $3.96 \; (0.14)$   & $3.25 \; (0.14)$   & $3.95 \; (0.14)$   \\
Distance           & $-0.08 \; (0.09)$  & $-0.24 \; (0.09)$  & $-1.68 \; (2.49)$  & $-5.93 \; (2.47)$  \\
Abs. Diff. Polity  & $-0.00 \; (0.01)$  & $-0.00 \; (0.01)$  & $-0.00 \; (0.01)$  & $-0.00 \; (0.01)$  \\
Alliance           & $0.37 \; (0.20)$   & $0.21 \; (0.21)$   & $0.40 \; (0.20)$   & $0.24 \; (0.21)$   \\
log-GDP (Sender)   & $0.30 \; (0.05)$   & $0.59 \; (0.04)$   & $0.30 \; (0.05)$   & $0.59 \; (0.04)$   \\
log-GDP (Receiver) & $0.17 \; (0.05)$   & $0.35 \; (0.04)$   & $0.16 \; (0.05)$   & $0.35 \; (0.04)$   \\
Mutual             & $-0.34 \; (0.43)$  &                    & $-0.33 \; (0.44)$  &                    \\
GWIDEG             & $-1.48 \; (0.31)$  &                    & $-1.49 \; (0.30)$  &                    \\
GWODEG             & $-2.84 \; (0.30)$  &                    & $-2.85 \; (0.30)$  &                    \\
GWOTP              & $-0.15 \; (0.11)$  &                    & $-0.15 \; (0.11)$  &                    \\
GWISP              & $0.22 \; (0.09)$   &                    & $0.22 \; (0.09)$   &                    \\
\hline
AIC                & $1772.34$          & $1891.65$          & $1773.36$          & $1893.14$          \\
BIC                & $1868.67$          & $1947.84$          & $1869.69$          & $1949.34$          \\
Log Likelihood     & $-874.17$          & $-938.82$          & $-874.68$          & $-939.57$          \\
\hline
\end{tabular}
\caption{Estimated coefficients of ERGMs and corresponding logistic regression
models for the international arms trade network in 2018 with covariates lagged by 4 and 5 years, resepectively.}
\label{table:coefficients_sm}
\end{center}
\end{table}

\section{Further results of the application to the global foreign exchange activity network}

 \subsection{Variance and covariance parameters}
\label{sec:amen_varcov}

\begin{table}[t]
\begin{center}
\begin{tabular}{l r r}
\hline
 & Estimate & Standard deviation \\
\hline
$\sigma^2_a$                             & $0.557$ & $0.165$ \\
$\sigma^2_b$                      & $1.863$ & $0.860$   \\
$\sigma_{ab}$                    & $-0.350$ & $0.250$   \\
$\rho$  & $0.364$ & $0.319$         \\
\hline
\end{tabular}
\caption{Estimated variance and covariance parameters for the AMEN model fitted on the global foreign exchange network in 1900.}
\label{table:varcov}
\end{center}
\end{table}

In Section 4.3 of the paper, we showcased the AME model by fitting it to the historical network of global foreign exchange activity in 1900. In illustrating the results we, for brevity, focused on the main effects of the covariates included, reported in Table \ref{table:amen}. But the AME model also estimates several variance and covariance parameters, which also have a meaningful interpretation. The estimates for those parameters are reported in Table \ref{table:varcov}. The first two parameters, $\sigma^2_a$ and $\sigma^2_b$, represent the estimated variances of the additive sender and receiver effects, respectively. We can see that receiver effects are much more variable then sender effects. This makes intuitive sense given that there are a few currencies which are traded by a large number of countries, while many currencies aren't traded at all outside of their origin countries. The skewness in the distribution of incoming ties thus induces a relatively large variance in the receiver effects.

The coefficient in the third row, $\sigma_{ab}$, measures the (global) correlation between sender and receiver effects of the same node. In this case, we can see that there is a slight negative correlation between the two, meaning that countries that trade many foreign currencies within their financial hub do not necessarily tend to have their home currency traded in many countries. Finally, the coefficient $\rho$ indicates the covariance between the residuals on the same node-pair, $\epsilon_{ij}$ and $\epsilon_{ji}$. This parameter quantifies the tendency towards reciprocity in the network. In this case, we can see that there is a slight positive tendency for ties to be reciprocated.

\subsection{Model diagnostics and goodness of fit}
\label{sec:annex_amen}

\begin{figure}[]
\centering 
         \includegraphics[width=0.9\linewidth]{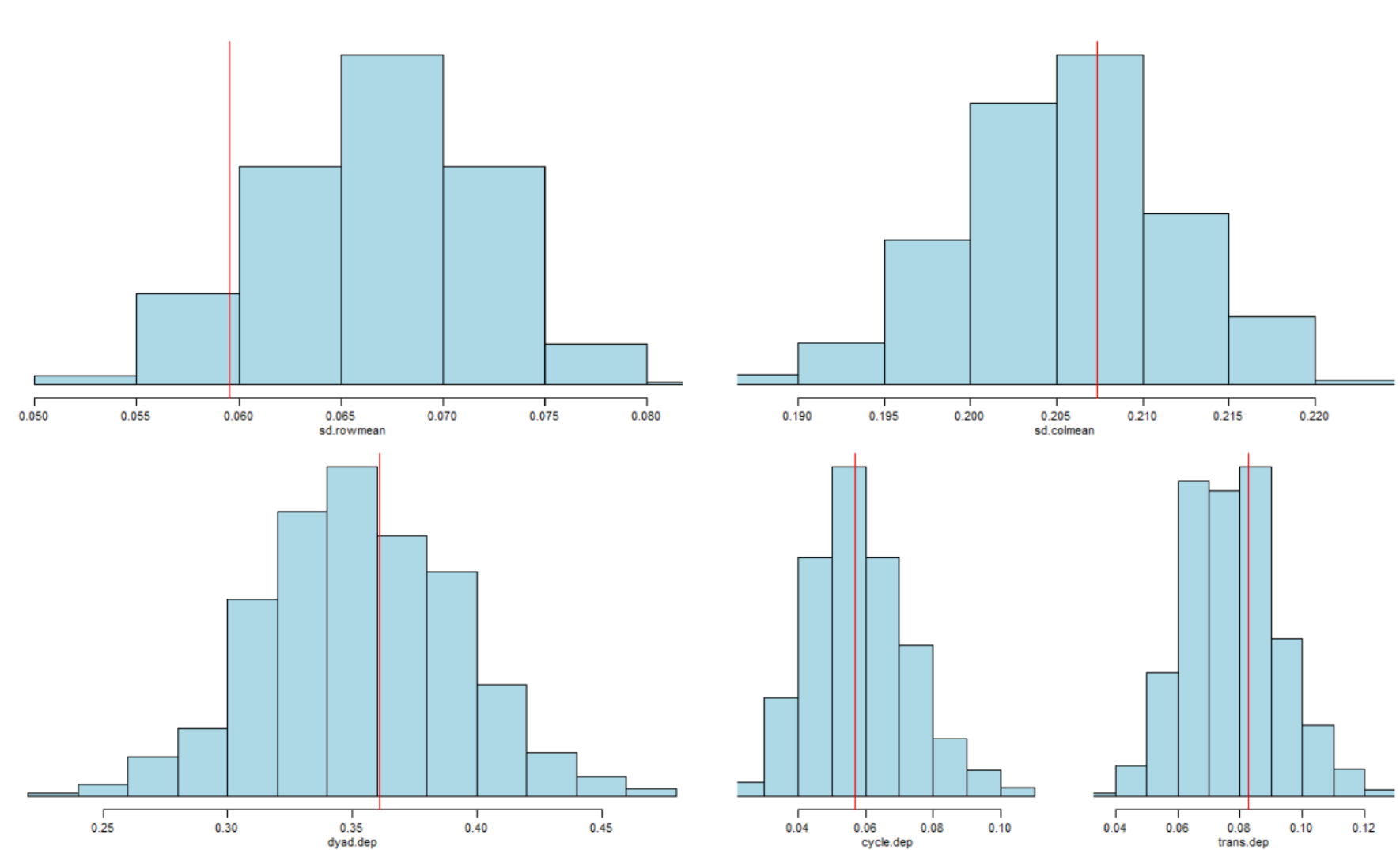}
\caption{Goodness of fit plots for the AME model applied to the global foreign exchange network in 1900. The model does an overall good job in replicating the observed network statistics.}\label{fig:amen_gof}
\end{figure}

Similarly as for ERGM, the goodness of fit for AME models is evaluated by comparing the statistics observed in the real world network with the distribution of the same statistics calculated for networks simulated from the model. Figure \ref{fig:amen_gof} depicts this comparison for first order effects (top panel) and for second- and third-order dependencies (bottom panel), as done by default in the \texttt{R} package \texttt{amen}  (see \citealp{hoff2015} for details). All in all, we can see that the model does a reasonably good job in preserving the network statistics in question. 

\begin{figure}[]
\centering 
         \includegraphics[width=0.9\linewidth]{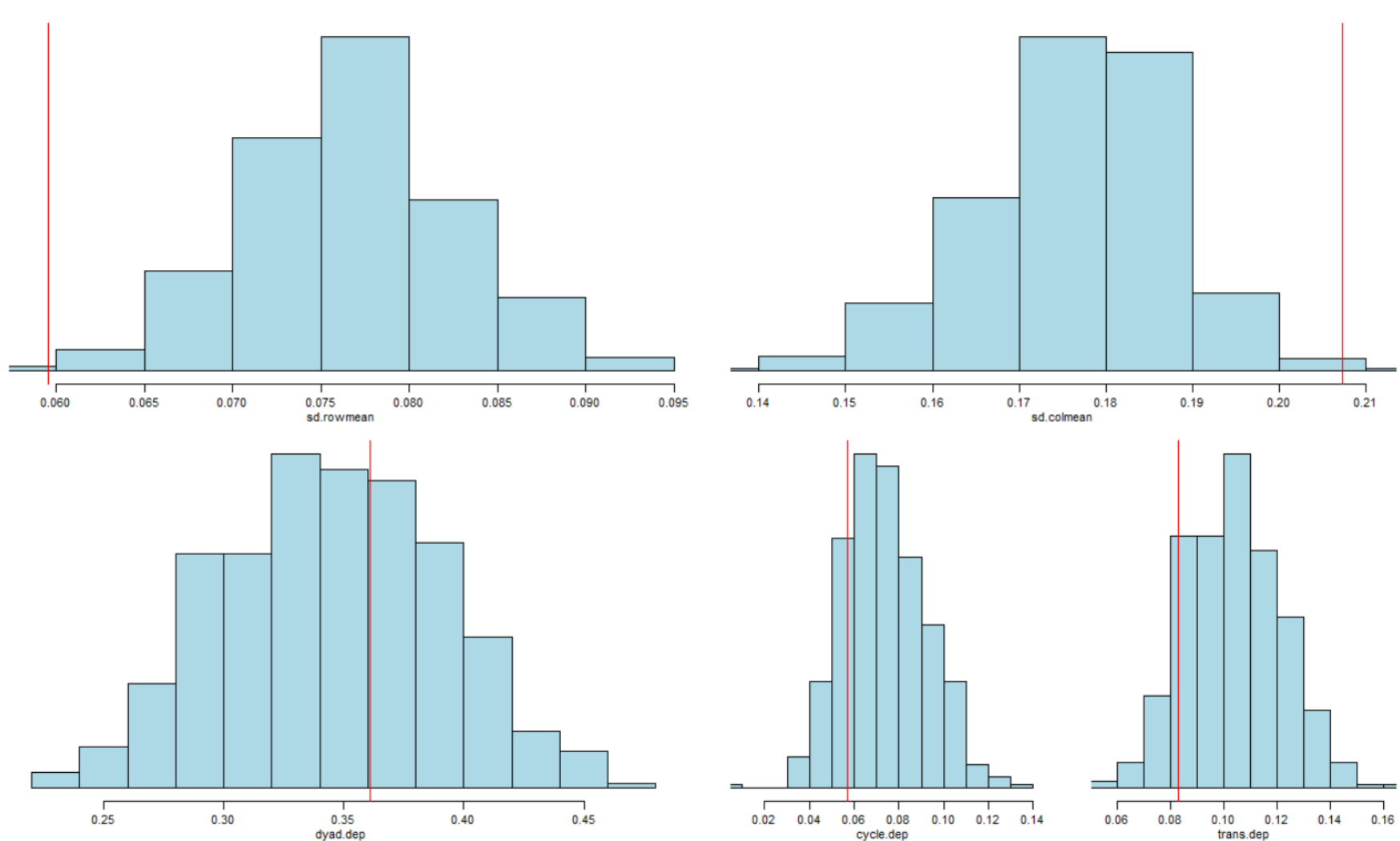}
\caption{Goodness of fit plots for the classical probit model applied to the global foreign exchange network in 1900. The model performs visibly worse than AME in replicating the observed network statistics.}\label{fig:probit_gof}
\end{figure}

We can also compare the goodness of fit of the AME model with that of the Probit model including the same exogenous covariates, but of course omitting the latent variables. Figure \ref{fig:probit_gof} depicts the same comparison of observed with simulated statistics just described, but for Probit instead of AME. From the plots therein we can see how the Probit model does a markedly worse job than the AME in reproducing first and third order dependencies, thus demonstrating an overall worse performance in capturing the mechanisms at play in the network. 

\begin{figure}[h]
\centering 
\includegraphics[width=0.9\linewidth]{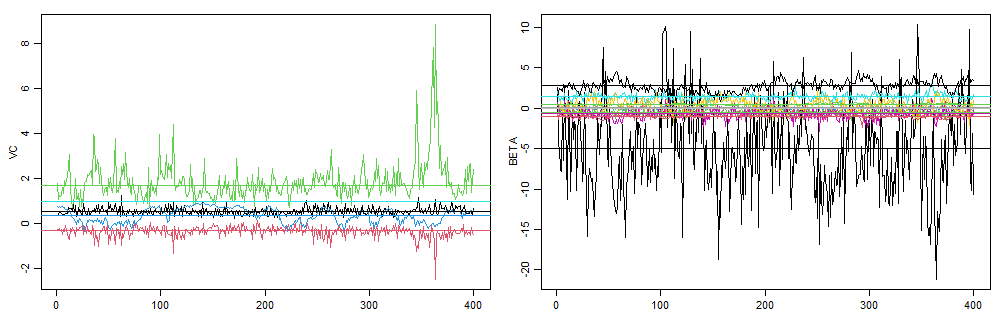}

\caption{MCMC diagnostics for the AME model applied to the global foreign exchange network in 1900. Visual inspection gives us confidence that the model has reasonably converged.}\label{fig:amen_mcmc}
\end{figure}

In Figure \ref{fig:amen_mcmc}, also produced by default by the \texttt{amen} package, we can further check how the coefficients and their variance vary across the MCMC iterations. In general, the fit is considered to be acceptable if no visible trends emerge in the chains. If, to the contrary, trends in the estimates are visible, the researcher must consider running the chain for more iterations and/or using alternative model formulations. In the case of Figure \ref{fig:amen_mcmc}, visual inspection gives us confidence that the AME model has reasonably converged.

\bibliographystylesupp{chicago}
\bibliographysupp{references}


\end{document}